\documentclass[a4paper,nobibnotes,nofootinbib]{revtex4}
\usepackage{amssymb}
\usepackage{amsmath}
\usepackage{epsfig}
\newcommand{\be}{\begin{equation}}
\newcommand{\ee}{\end{equation}}
\newcommand{\bea}{\begin{eqnarray}}
\newcommand{\eea}{\end{eqnarray}}

\newcommand{\D}{\displaystyle}
\newcommand{\g}{\gamma}
\newcommand{\f}{\frac}

\newcommand{\intc}[1]{{\int\frac{d#1}{2i\pi}}}
\newcommand\lr[1]{{\left({#1}\right)}}
\begin{document}
\title{Small-x QCD effects in forward-jet and Mueller-Navelet jet production}
\author{C. Marquet}\email{marquet@spht.saclay.cea.fr}
\affiliation{Service de physique th{\'e}orique, CEA/Saclay, 91191 Gif-sur-Yvette 
cedex, France\\URA 2306, unit{\'e} de recherche associ{\'e}e au CNRS}
\author{C. Royon}\email{royon@hep.saclay.cea.fr}
\affiliation{DAPNIA/Service de physique des particules, CEA/Saclay, 91191 
Gif-sur-Yvette cedex, France\\and Fermi National Accelerator Laboratory, 
Batavia, Illinois 60510, USA}
%%%%%%%%%%%%%%%%%%%%%%%%%%%%%%%%%%%%%%%%%%%%%%%%%%%%%%%%%%%%%%%%%%
%%%%%%%%%%%%%%%%%%%%%%%%   Abstract   %%%%%%%%%%%%%%%%%%%%%%%%%%%%
%%%%%%%%%%%%%%%%%%%%%%%%%%%%%%%%%%%%%%%%%%%%%%%%%%%%%%%%%%%%%%%%%%
\begin{abstract}

We investigate small$-x$ QCD effects in forward-jet production in deep inelastic 
scattering in the kinematic regime where the virtuality of the photon and the 
transverse momentum of the jet are two hard scales of about the same magnitude.
We show that the data from HERA published by the H1 and ZEUS collaborations are 
well described by leading-logarithmic BFKL predictions. Parametrizations 
containing saturation effects expected to be relevant at higher energies also 
compare well to the present data. We extend our analysis to Mueller-Navelet jets 
at the LHC and discuss to what extent this observable could test these 
small$-x$ effects and help distinguishing between the different descriptions.

\end{abstract}
\maketitle
%%%%%%%%%%%%%%%%%%%%%%%%%%%%%%%%%%%%%%%%%%%%%%%%%%%%%%%%%%%%%%%%%%
%%%%%%%%%%%%%%%%%%%%%%   Introduction   %%%%%%%%%%%%%%%%%%%%%%%%%%
%%%%%%%%%%%%%%%%%%%%%%%%%%%%%%%%%%%%%%%%%%%%%%%%%%%%%%%%%%%%%%%%%%
\section{Introduction}
\label{1}

The Regge limit of perturbative QCD comes about when the centre-of-mass energy 
in a collision is much bigger than the fixed hard scales of the problem. In this 
limit usually called the small$-x$ regime, parton densities inside the 
projectiles grow with increasing energy, leading to the growth of the scattering 
amplitudes. As long as the densities are not too high, this growth is described 
by the Balitsky-Fadin-Kuraev-Lipatov (BFKL) equation~\cite{bfkl} that resums the 
leading logarithms. As the parton density becomes higher and the scattering 
amplitudes approach the unitarity limit, one enters a regime called 
saturation~\cite{glr,glr+,mv,jimwlk} where the BFKL evolution breaks down and 
parton densities saturate.

In the past years, as colliders started to explore the small$-x$ regime, 
proposals were made to test the relevance of the BFKL equation at the available 
energies. In this paper we concentrate on two of the proposed measurements: 
forward jets~\cite{dis} in deep inelastic scattering (DIS) and Mueller-Navelet 
jets~\cite{mnj} in hadron-hadron collisions. Forward-jet production is a process 
in which the virtual photon interacts with the proton and a jet is detected in 
the forward direction of the proton. The virtuality of the photon and the 
squared transverse momentum of the jet are hard scales of about the same 
magnitude. In the case of Mueller-Navelet jets, a proton interacts with 
another proton or antiproton and a jet is detected in each of the two forward 
directions; the transverse momenta of the jets are as well hard scales of about 
the same magnitude. If the total energy in the photon-proton (for forward jets) 
or proton-proton (for Mueller-Navelet jets) collision is large enough, these 
processes feature the kinematics corresponding to the Regge limit. 

The description of forward jets with fixed-order perturbative QCD in the Bjorken 
limit amounts in the following. Large logarithms coming from the strong ordering 
between the soft proton scale and the hard forward-jet scale are resummed using 
the Dokshitzer-Gribov-Lipatov-Altarelli-Parisi (DGLAP) evolution 
equation~\cite{dglap} and the hard cross-section is computed at fixed order in 
the coupling constant. In the small$-x$ regime, due to the extra ordering between 
the total energy and the hard scales, other large logarithms arise and should be 
resummed within the hard cross-section itself. In other words, the inclusion of 
small$-x$ effects aims at improving QCD predictions by replacing fixed-order hard 
cross-sections with resummed hard cross-section, using the BFKL equation or, at 
even higher energies, using resummations that include saturation effects.

To study different observables in this small$-x$ regime, a convenient approach 
is to formulate the cross-sections in terms of scattering amplitudes for 
colorless combinations of partons. The simplest of those is the $q\bar q$ 
dipole, a quark-antiquark pair in the color singlet state; it describes for 
instance the interaction of a virtual photon. Any colorless $gg,$ $q\bar q 
g,...$ multiplets can a priori be involved, for instance the gluon-gluon $(gg)$ 
dipole is what describes gluon emissions. 

To compute the evolution of those scattering amplitudes with energy, the QCD 
dipole model~\cite{mueller} has been developed. This formalism constructs the 
light-cone wavefunction of a $q\bar q$ dipole in the leading logarithmic 
approximation. As the energy increases, the original dipole evolves and the 
wavefunction of this evolved dipole is described as a system of elementary 
$q\bar q$ dipoles. When this system of dipoles scatters on a target, the 
scattering amplitude has been shown to obey the BFKL equation. Interestingly 
enough, the dipole formalism was shown to be also well-suited to include density 
effects and non-linearities that lead to saturation and unitarization of the 
scattering amplitudes~\cite{bk,edal}. This is why the dipole picture is suitable 
for investigating the small$-x$ regime of QCD, it allows to study both BFKL and 
saturation effects within the same theoretical framework. 

The formulation of the forward-jet and Mueller-Navelet jet processes in terms of 
dipole amplitudes has been addressed in~\cite{marq}. In both cases the problem 
is similar to the one of onium-onium scattering: the growth with energy of the 
total cross-section due to BFKL evolution is damped by saturation effects 
which arise purely perturbatively. For instance in the large$-N_c$ limit, that 
involves multiple Pomeron exchanges. In our study, we consider both the BFKL 
energy regime and saturation effects. We shall implement saturation in a very 
simple way, using a phenomenological parametrization inspired by the 
Golec-Biernat and W\"usthoff (GBW) approach~\cite{golec} which gave a good 
description of the proton structure functions with a very few parameters.

First comparisons of small$-x$ predictions with forward-jet data from HERA were 
quite successful: the first sets of data published by the H1~\cite{h199} and 
ZEUS~\cite{zeus99} collaborations are well described by the leading-logarithmic 
(LL) BFKL predictions~\cite{fjets} and also show compatibility with saturation 
parametrizations~\cite{mpr,dis05}. In both cases, the DGLAP resummation 
associated accounted for leading logarithms.

Very recently, new forward-jet experimental results have been 
published~\cite{zeusnew,h1new}. They involve a broader range of observables with 
several differential cross-sections and go to smaller values of $x$ than the 
previous measurements. As QCD at next-to-leading order (NLO) is not sufficient to 
describe the small$-x$ data, we shall address the following issues: whether the 
BFKL-LL predictions keep being in good agreement and whether saturation 
parametrizations still show compatibility. The first part of the present work is 
devoted to those questions.

In the second part of the paper, we deal with Mueller-Navelet jets. We display  
BFKL-LL predictions in the LHC energy range for different differential 
cross-sections. We compare them with saturation predictions obtained from our 
parametrizations of saturation effects constrained by the forward-jet data. We 
propose different measurements and discuss their potential for identifying BFKL 
and saturation behaviors.

The paper will be organized as follows. In section II, we compute the 
forward-jet cross-section in the high-energy regime and express it in terms of 
a dipole-dipole cross-section. We compare the BFKL-LL predictions and the
saturation parametrization with the new H1 and ZEUS data for several 
differential cross-sections. In section II, we compute the Mueller-Navelet 
jet cross-section and show the BFKL and saturation predictions for LHC energies 
and for several differential cross-sections. The final section V is devoted to 
conclusion and outlook.

%%%%%%%%%%%%%%%%%%%%%%%%%%%%%%%%%%%%%%%%%%%%%%%%%%%%%%%%%%%%%%%%%%
%%%%%%%%%%%%%%%%%%%%%%%%   Section 1   %%%%%%%%%%%%%%%%%%%%%%%%%%%
%%%%%%%%%%%%%%%%%%%%%%%%%%%%%%%%%%%%%%%%%%%%%%%%%%%%%%%%%%%%%%%%%%
\section{Forward-jet production}
\label{2}

Forward-jet production in a lepton-proton collision is represented in Fig.1 with 
the different kinematic variables. We denote $\sqrt{s}$ the total energy of the 
lepton-proton collision and $Q^2$ the virtuality of the intermediate photon that 
undergoes the hadronic interaction. We shall use the usual kinematic variables 
of deep inelastic scattering: $x\!=\!Q^2/(Q^2\!+\!W^2)$ and $y\!=\!Q^2/(xs)$ 
where $W$ is the center-of-mass energy of the photon-proton collision. In 
addition, $k_T\!\gg\!\Lambda_{QCD}$ is the jet transverse momentum and $x_J$ its 
longitudinal momentum fraction with respect to the proton. In the following, we 
compute the forward-jet cross-section in the high-energy limit, recall the BFKL 
predictions and give our formulation of the saturation model.

\subsection{Formulation}

The QCD cross-section for forward-jet production reads
\be
\f{d^{(4)}\sigma}{dxdQ^2dx_Jdk_T^2}=\f{\alpha_{em}}{\pi xQ^2}
\left\{\lr{\f{d\sigma^{\g*p\!\rightarrow\!JX}_T}{dx_Jdk_T^2}
+\f{d\sigma^{\g*p\!\rightarrow\!JX}_L}{dx_Jdk_T^2}}(1-y)
+\f{d\sigma^{\g*p\!\rightarrow\!JX}_T}{dx_Jdk_T^2}\f{y^2}2\right\}\ 
,\label{one}\ee
where $d\sigma^{\g*p\!\rightarrow\!JX}_{T,L}/dx_Jdk_T^2$ is the cross-section 
for forward-jet production in the collision of the transversely (T) or 
longitudinally (L) polarized virtual photon with the target proton.

We now consider the high-energy regime $x\!\ll\!1.$ In an appropriate frame 
called the dipole frame, the virtual photon undergoes the hadronic interaction 
via a fluctuation into a dipole. The dipole then interacts with the target 
proton and one has the following factorization
\be
\f{d\sigma^{\g*p\!\rightarrow\!JX}_{T,L}}{dx_Jdk_T^2}=\int d^2 r 
\int_0^1 dz\ |\psi_{T,L}^{\g}(r,z;Q)|^2
\f{d\sigma_{q\bar q}}{dx_Jdk_T^2}(r)\ .\label{fact}\ee
The wavefunctions $\psi^\gamma_T$ and $\psi^\gamma_L$ describe the splitting of 
the photon on the dipole and $d\sigma_{q\bar q}/dx_Jdk_T^2$ is the cross-section
for forward-jet production in the dipole-proton collision. $\psi^\gamma_T$ and 
$\psi^\gamma_L$ are given by
\be
|\psi^{\g}_T(r,z;Q)|^2=
\frac{\alpha_{em}N_c}{2\pi^2}\sum_f e_f^2
(z^2+(1-z)^2)z(1\!-\!z)Q^2K_1^2\lr{\sqrt{z(1\!-\!z)}Q|r|}
\label{vpwfT}\ee
\be
|\psi^{\g}_L(r,z;Q)|^2=
\frac{\alpha_{em}N_c}{2\pi^2}\sum_f e_f^2
4Q^2 z^2(1-z)^2 K_0^2\lr{\sqrt{z(1\!-\!z)}Q|r|}
\label{vpwfL}\ee
for a transversely (\ref{vpwfT}) and longitudinally (\ref{vpwfL}) polarized 
photon where $e_f$ is the charge of the quark\footnote{We consider massless 
quarks and sum over four flavors in (\ref{vpwfT}) and (\ref{vpwfL}). This is 
justified considering the rather high values of the photon virtuality 
($Q^2\!>\!5\ \mbox{GeV}^2$) used for the measurement.} with flavor $f.$ The 
integration variable $r$ is the transverse size of the $q\bar q$ pair and $z$ is 
the longitudinal momentum fraction of the antiquark with respect to the photon. 
In the leading logarithmic approximation we are interested in, the cross-section 
$d\sigma_{q\bar q}/dx_Jdk_T^2$ does not depend on $z$ but only on the dipole 
size $r.$ This cross-section has been computed in~\cite{marq} where it was shown 
that the emission of the forward jet can be described through the interaction of 
an effective gluonic $(gg)$ dipole:
\begin{figure}[t]
\begin{center}
\epsfig{file=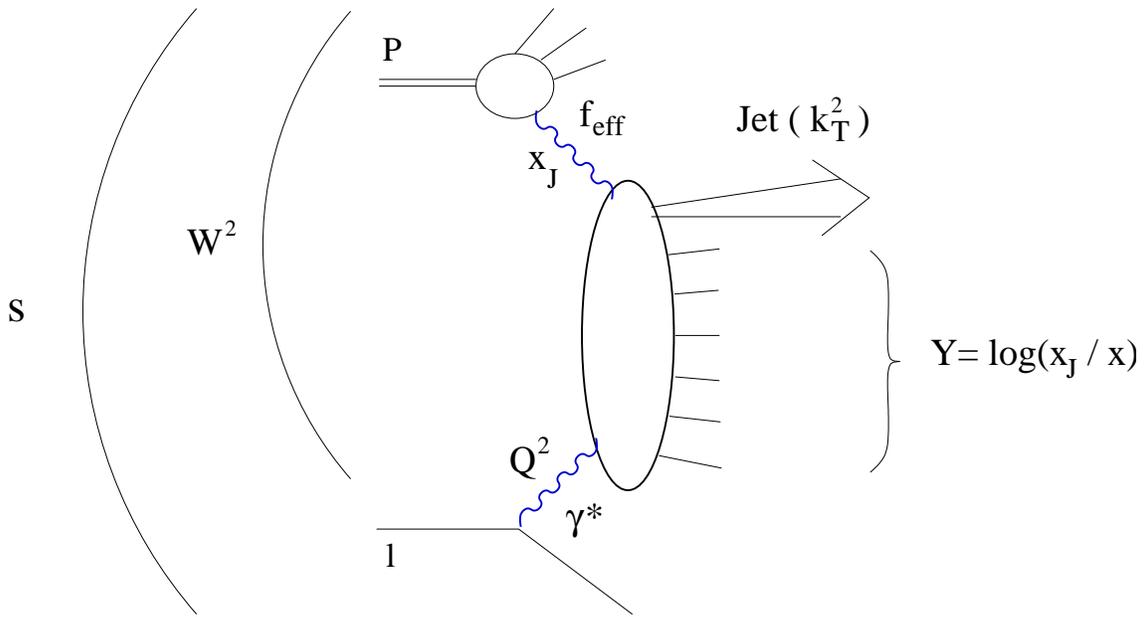,width=15cm}
\caption{Production of a forward jet in a proton-lepton collision. The kinematic 
variables of the problem are displayed. $Q^2$ is the virtuality of the photon 
that undergoes the hadronic interaction. $s$ and $W^2$ are the total energies 
squared in the lepton-proton and photon-proton collisions respectively. $k_T$ is 
the transverse momentum of the forward jet and $x_J$ is its longitudinal 
momentum fraction with respect to the incident proton. $Y$ is the rapidity 
interval between the two hard probes.}
\end{center}
\label{F1}
\end{figure} 
\be
\f{d\sigma_{q\bar q}}{dx_Jdk_T^2}(r)=\f{\pi N_c}{16k_T^2}f_{eff}(x_J,k_T^2)
\int_0^\infty d\bar{r}\ J_0(k_T\bar{r})\ \f{\partial}{\partial 
\bar{r}}\lr{\bar{r}
\f{\partial}{\partial \bar{r}}\ \sigma_{(q\bar q)(gg)}(r,\bar{r},Y)}
\label{fjdcs}\ee
with $Y\!=\!\log(x_J\!/\!x)$ the rapidity assumed to be very large. 
$\sigma_{(q\bar q)(gg)}(r,\bar{r},Y)$ is the cross-section in the collision of a 
$q\bar q$ dipole of size $r$ with a $gg$ dipole of size $\bar{r}$ with total 
rapidity $Y.$ $f_{eff}(x_J,k_T^2)$ is the effective parton distribution function 
and resums the leading logarithms $\log(k_T^2/\Lambda_{QCD}^2).$ It has the 
following expression
\be
f_{eff}(x_J,k_T^2)=g(x_J,k_T^2)+
\f{C_F}{N_c}\lr{q(x_J,k_T^2)+\bar{q}(x_J,k_T^2)}\ ,
\label{sf}\ee
where $g$ (resp. $q$, $\bar{q}$) is the gluon (resp. quark, antiquark) 
distribution function in the incident proton. 

Let us comment formula (\ref{fjdcs}). Since the forward jet measurement involves 
perturbative values of $k_T$ and moderate values of $x_J,$ it is not surprising 
that formula (\ref{fjdcs}) features the collinear factorization of $f_{eff};$ 
note also that $k_T^2$ has been chosen as the factorization scale. The remaining 
hard interaction is between a $gg$ dipole and the incident $q\bar q$ dipole of 
size $r.$ The $gg$ dipole emerges as the effective degree of freedom for the 
gluon emission at high energies~\cite{marq}. This feature has been pointed out 
several times~\cite{ggdip}. 

Formulae (\ref{one})-(\ref{sf}) express the forward-jet observable 
(\ref{one}) in terms of the cross-section $\sigma_{(q\bar q)(gg)}$ which 
contains the high-energy QCD dynamics: the problem is analogous to the one of 
onium-onium scattering. In the next subsection (II-B), we deal with the BFKL 
energy regime for which the interaction between the $q\bar q$ dipole and $gg$ 
dipole is restricted to a Pomeron exchange. In that case of course, our 
formulation is equivalent to the $k_T-$factorization approach. In subsection 
(II-C), we go beyond $k_T-$factorization and investigate the saturation regime 
in which $\sigma_{(q\bar q)(gg)}$ a priori contains any number of gluon 
exchanges.

\subsection{The BFKL energy regime}

The BFKL $q\bar q$-$gg$ dipole-dipole cross-section reads (see for 
instance~\cite{samuel})
\be
\sigma^{BFKL}_{(q\bar q)(gg)}(r,\bar{r},Y)=2\pi\alpha_s^2r^2
\intc{\g}\f{\lr{\bar{r}/r}^{2\g}}{\g^2(1\!-\!\g)^2}\
\exp{\lr{\D\f{\alpha_sN_c}{\pi}\chi(\g)Y}}
\label{sdd}
\ee
with the complex integral running along the 
imaginary axis from $1/2\!-\!i\infty$ to $1/2\!+\!i\infty$ and
with the BFKL kernel given by
\be
\chi(\g)=2\psi(1)-\psi(1-\g)-\psi(\g)
\ee
where $\psi(\g)$ is the logarithmic derivative of the Gamma function.
It comes about when the interaction between the $q\bar q-$dipole and the 
$gg-$dipole is restricted to a two-gluon exchange. Summing the 
leading-logarithmic contributions of ladders with any number of real gluon 
emissions, one obtains the BFKL Pomeron and 
the resulting growth of the cross-section with rapidity. 

Inserting (\ref{sdd}) in (\ref{fjdcs}) and (\ref{fact}), one obtains
\be
\f{d\sigma^{\g*p\!\rightarrow\!JX}_{T,L}}{dx_Jdk_T^2}=
\f{\pi^2N_c\alpha_s^2}{4k_T^2Q^2}f_{eff}(x_J,k_T^2)\intc{\g}
\lr{\f{Q^2}{k_T^2}}^\g\f{4^\g\Gamma(\g)\ \phi^\g_{T,L}(\g)}
{(1\!-\!\g)\ \Gamma(2\!-\!\g)}\exp{\lr{\D\f{\alpha_sN_c}{\pi}\chi(\g)Y}}
\label{bfklres}\ee
where we have defined the following Mellin-transforms 
\be
\phi^\g_{T,L}(\g)=\int d^2r(r^2Q^2)^{1-\g}\int_0^1 dz|\psi_{T,L}^{\g}(r,z;Q)|^2
\ee
which are given by
\be
\lr{\begin{array}{cc}
\phi^\g_{T}(\g)\\\phi^\g_{L}(\g)
\end{array}}
=\f{2\alpha_{em}N_c}{\pi}\sum_q e_q^2\f{1}{4^\g\g}
\f{\Gamma^2(1+\g)\Gamma^2(1-\g)\Gamma^2(2-\g)}
{\Gamma(2-2\g)\Gamma(2+2\g)(3-2\g)}
\lr{\begin{array}{cc}(1+\g)(2-\g)\\2\g(1-\g)\end{array}}\ .
\label{phig}
\ee
Inserting formula (\ref{bfklres}) into (\ref{one}) gives the forward-jet 
cross-section in the BFKL energy regime. One can easily show that the result is 
identical to the one obtained using $k_T-$factorization~\cite{fjets,ktfac}. The 
only undetermined parameter is $\bar\alpha\!\equiv\!\alpha_sN_c/\pi$ (with 
$\alpha_s$ the strong coupling constant kept fixed) which appears in the 
exponential in formula (\ref{bfklres}).

\subsection{The saturation regime}

Contrary to the BFKL case, the onium-onium cross-section in the saturation 
regime has not yet been computed from QCD. Studies are being carried out to 
identify the dominant terms in the multiple gluon 
exchanges~\cite{edal,poml1,poml2,poml3} but 
the cross-section $\sigma^{sat}_{(q\bar q)(gg)}$ remains unknown. To take into 
account saturation effects, we are led to use a phenomenological 
parametrization. We consider the following model introduced in~\cite{mpr} which 
is inspired by the GBW approach:
\be
\sigma^{sat}_{(q\bar q)(gg)}(r,\bar{r},Y)=4\pi\alpha_s^2\sigma_0
\lr{1-\exp\lr{-\f{r_{\rm eff}^2(r,\bar{r})}{4R_0^2(Y)}}}\ .
\label{sigmadd}
\ee
The dipole-dipole {\it effective} radius $r^2_{\rm  eff}(r,\bar{r})$ is defined 
through the two-gluon exchange:
\be 
4\pi\alpha_s^2r^2_{\rm eff}(r,\bar{r})\equiv
\sigma^{BFKL}_{(q\bar q)(gg)}(r,\bar{r},0)
=4\pi\alpha_s^2\min(r^2\!,\bar{r}^2)\left\{1\!+\!\log
\frac{\max(r\!,\bar{r})}{\min(r\!,\bar{r})}\right\}\label{reff}\ .\ee
For the saturation radius we use the parametrization
$R_0(Y)\!=\!e^{-\f{\lambda}2\left(Y-Y_0\right)}/Q_0$ with $Q_0\!\equiv\!1\ GeV.$

Let us express the cross-section in terms of a double Mellin-transform:
\be
\sigma^{sat}_{(q\bar q)(gg)}(r,\bar{r},Y)=4\pi\alpha_s^2\sigma_0
\intc{\g}\intc{\tau}\lr{\f{r^2}{4R_0^2(Y)}}^{1-\g}
\lr{\f{\bar{r}^2}{4R_0^2(Y)}}^{\tau}g(\g,\tau)\label{dmt}
\ee
with
\bea
g(\g,\tau)=\int_0^\infty du^2\int_0^\infty d\bar{u}^2 u^{2\g-4}
\bar{u}^{-2\tau-2}\lr{1-e^{-r_{eff}^2(u,\bar{u})}}=
\f{2\Gamma{(\g-\tau)}}{1+\tau-\g}\ 
\left\{\Psi(1,3\!+\!\tau\!-\!\g,2\tau)+
\Psi(1,3\!+\!\tau\!-\!\g,2\!-\!2\g)\right\}\nonumber 
\label{Psi}
\\ 0<Re(\tau),\;Re(\g),\;Re(\g-\tau)<1  \hspace{12cm}
\eea
where the confluent hypergeometric function of Tricomi $\Psi(1,a,b)$ can be 
expressed~\cite{prud} in terms of incomplete Gamma functions. 
Inserting (\ref{dmt}) in (\ref{fjdcs}) and (\ref{fact}), one obtains
\be
\f{d\sigma^{\g*p\!\rightarrow\!JX}_{T,L}}{dx_Jdk_T^2}=
\f{\pi^2N_c\alpha_s^2\sigma_0}{8Q^2k_T^2R_0^2(Y)}f_{eff}(x_J,k_T^2)
\intc{\g}(4Q^2R_0^2(Y))^{\g}\phi^{\g}_{T,L}(\g)
\intc{\tau}(4k_T^2R_0^2(Y))^{-\tau}
\f{4^\tau\tau^2\Gamma(\tau)}{\Gamma(1\!-\!\tau)}\
g(\g,\tau)\ .\label{satres}\ee
Inserting formula (\ref{satres}) into (\ref{one}) gives our parametrization of 
the forward-jet cross-section in the saturation regime. The parameters are 
$\lambda,$ $Y_0$ and the normalization $\sigma_0.$

\subsection{Fixing the parameters}

The first sets of data published by the H1~\cite{h199} and ZEUS~\cite{zeus99} 
collaborations regarded the measurement of $d\sigma/dx.$ In previous 
studies, we fitted the BFKL-LL~\cite{fjets} and saturation 
parametrization~\cite{mnj} on those data with the cut $x\!<\!10^{-2}.$  Despite 
corresponding different energy regimes, in both cases we obtained good 
descriptions with $\chi^2$ values of about 1. The obtained values of the 
parameters and the $\chi^2$ of the fits are given in Table I.

\begin{table}[h]
\begin{center}
\begin{tabular}{|c||c|c|c|} \hline
 fit & parameters & $1/R_0(Y\!=\!0)$ & $\chi^2 (/\mbox{d.o.f.})$ \\
\hline\hline
 BFKL-LL       & $4\bar\alpha\log(2)\!=\!0.430$  &  ------  & 12 (/13)\\
 strong sat. & $\lambda=0.402$ and $Y_0=-0.82$ & 1.18 Gev & 6.8 (/11)\\
 weak sat.   & $\lambda=0.370$ and $Y_0=8.23$  & 0.22 Gev & 8.3 (/11)\\
\hline
\end{tabular}
\end{center}\caption{Results of the BFKL and saturation fits to the first HERA 
forward-jet data. The saturation fits shows two independent solutions showing 
either strong or weak saturation parameters (see text).}
\end{table}

In the BFKL-LL case, the only parameter is $\bar\alpha$ and the value obtained 
was $4\bar\alpha\log(2)\!=\!0.430.$ For the saturation fit, the two relevant 
parameters are $\lambda$ and $Y_0$ and the fit showed two $\chi^2$ minima for 
$(\lambda\!=\!0.402,Y_0\!=\!-0.82)$ and $(\lambda\!=\!0.370,Y_0\!=\!8.23).$ We 
shall refer to the first (resp. second) solution as a strong (resp. weak) 
saturation parametrization. Indeed, the first saturation minimum corresponds 
to strong saturation effects as, for typical values of $Y,$ the saturation 
scale $1/R_0$ is about 5 Gev which is the value of a typical $k_T.$ The second 
saturation minima corresponds to small saturation effects and rather describes 
BFKL physics.

Along with formulae (\ref{bfklres}) and (\ref{satres}), the values of the 
parameters given in Table I completely determine the BFKL-LL predictions and two 
parametrizations for the saturation model. We are now going to compare these 
with the very recent data without any adjustment of the parameters. This will 
provide a strong test of those small$-x$ effects.

\subsection{Comparison to the 2005 HERA data}

\begin{figure}[t]
\begin{minipage}[t]{88mm}
\includegraphics[width=8.6cm]{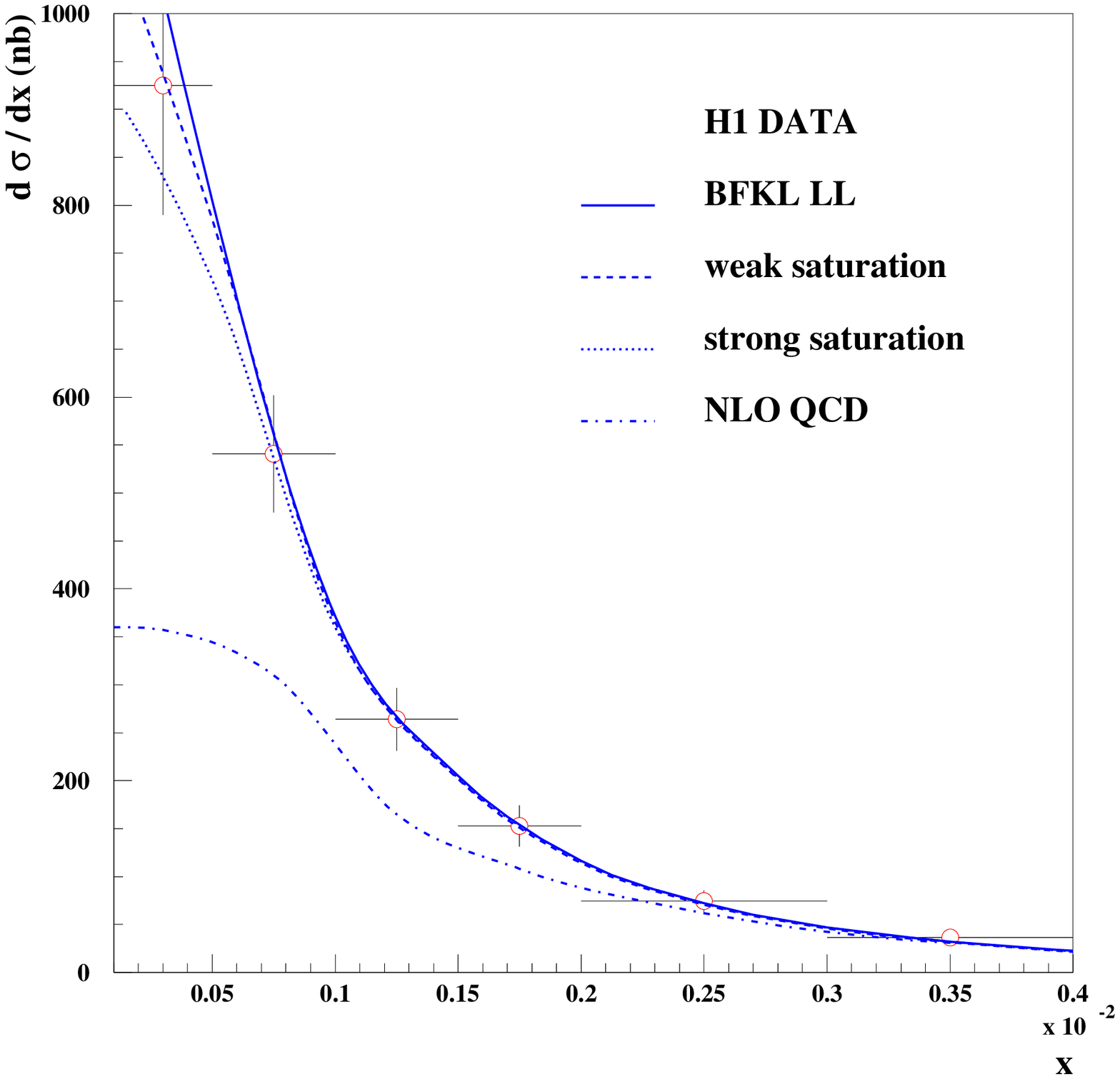}
\end{minipage}
\hspace{\fill}
\begin{minipage}[t]{88mm}
\includegraphics[width=8.6cm]{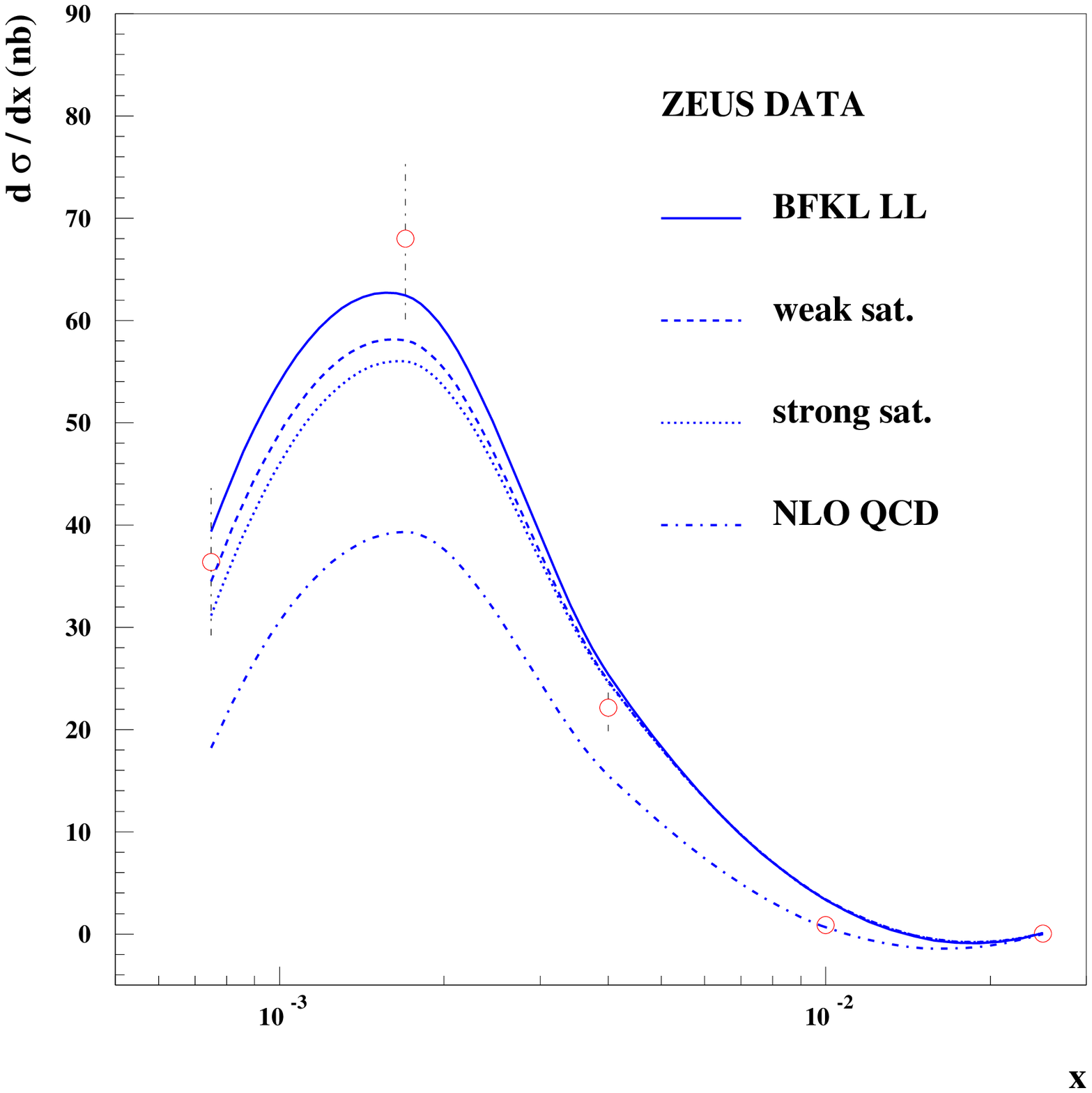}
\end{minipage}
\caption{The forward-jet cross-section $d\sigma/dx.$ The points are measurement 
by the H1 (left plot) and ZEUS (right plot) collaborations. The lines are 
comparisons with BFKL-LL predictions (full lines) and the two saturation 
parametrizations (dotted and dashed lines). In both cases, there is good 
agreement with the data. For comparison, fixed-order QCD predictions at NLO are 
also displayed.}
\label{F2}
\end{figure}

\begin{figure}[ht!]
\begin{minipage}[t]{88mm}
\includegraphics[width=8.6cm]{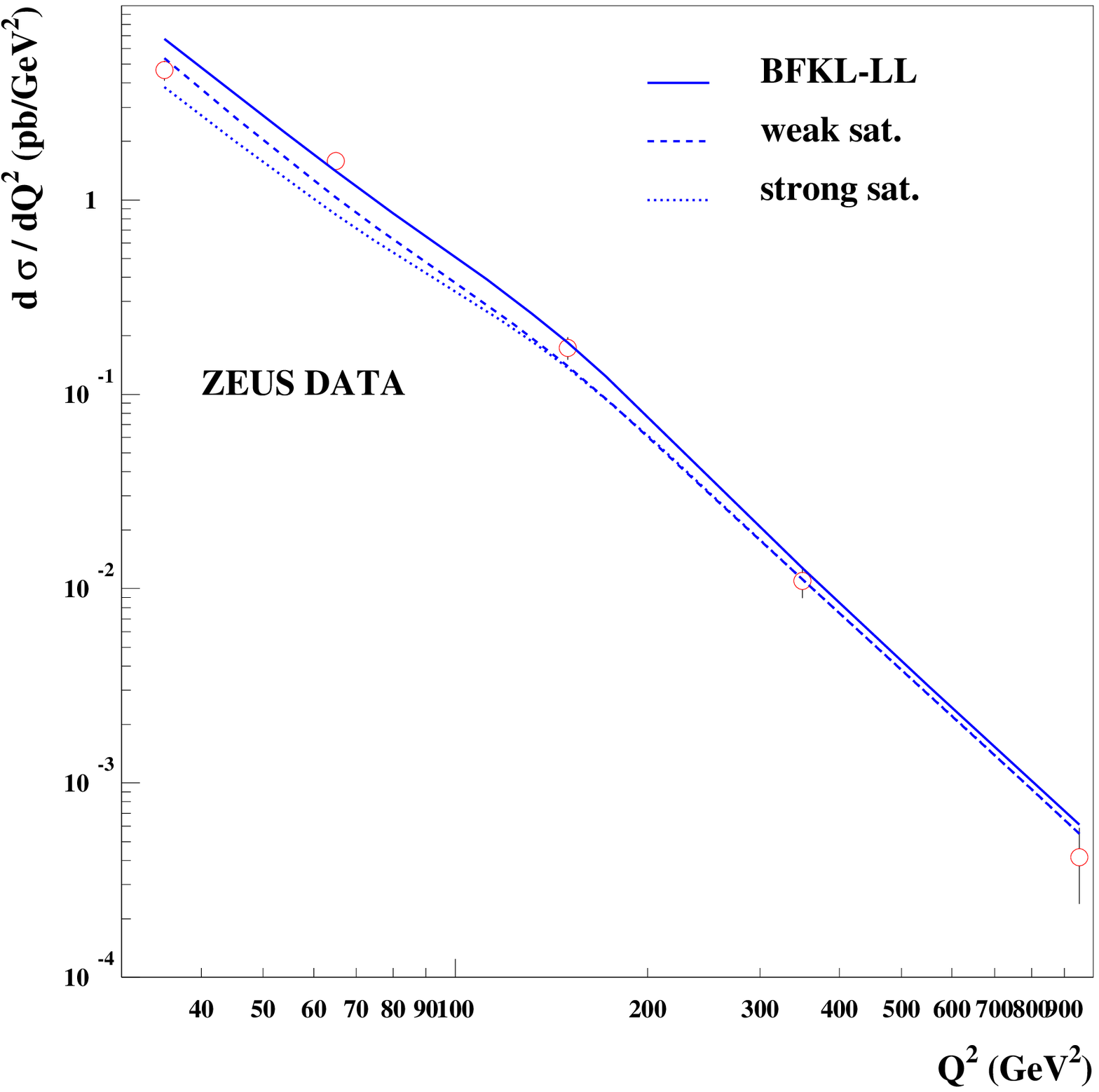}
\end{minipage}
\hspace{\fill}
\begin{minipage}[t]{88mm}
\includegraphics[width=8.6cm]{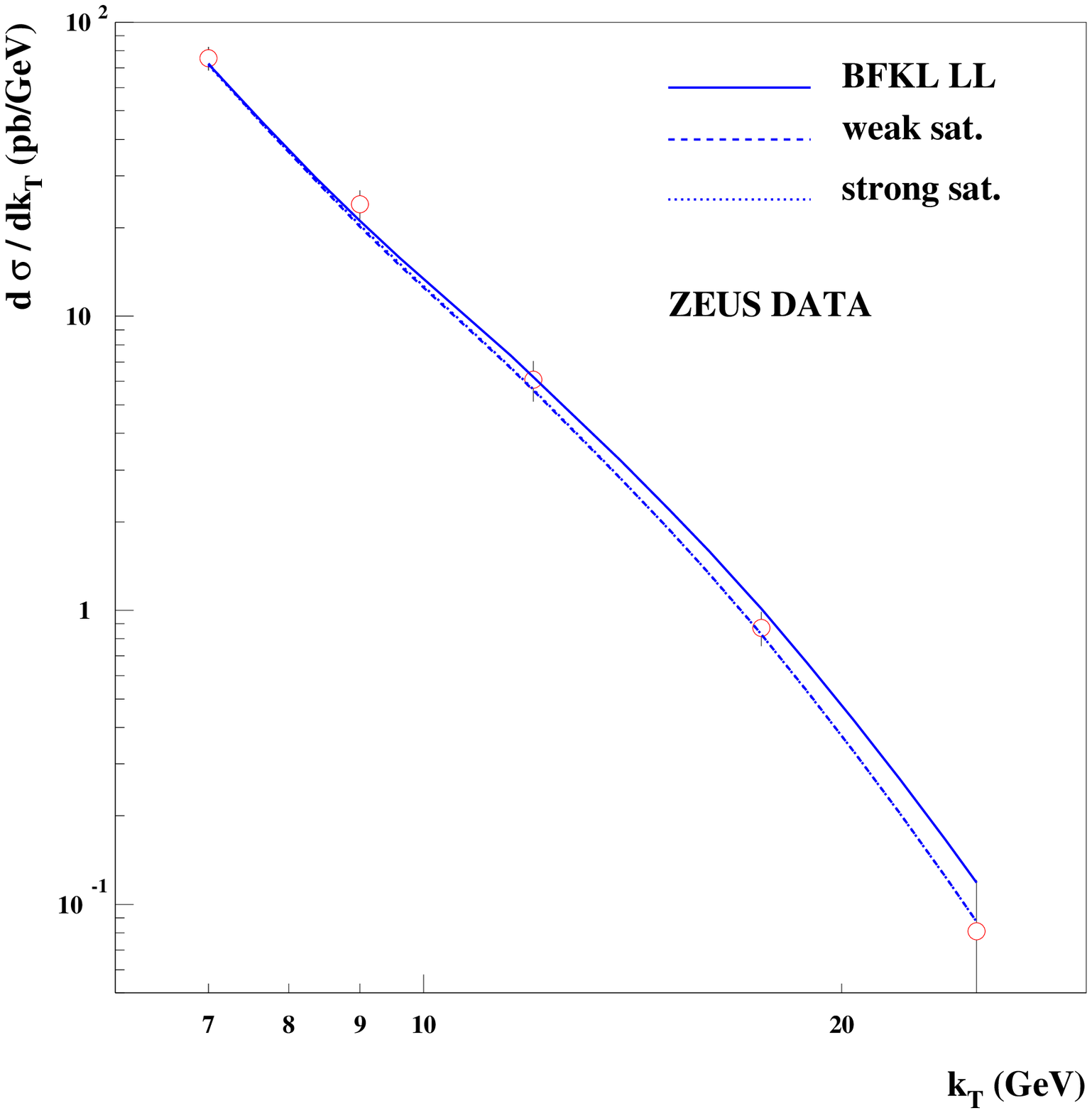}
\end{minipage}
\caption{The forward-jet cross-sections $d\sigma/dQ^2$ (left plot) and 
$d\sigma/dk_T$ (right plot). The points are measurement by the ZEUS 
collaboration. The lines are comparisons with BFKL-LL predictions (full lines) 
and the two saturation parametrizations (dotted and dashed lines). For the two 
observables, there is good agreement with the data.}
\label{F3}
\end{figure}

We want to compare the cross-section (\ref{one}) obtained from the BFKL-LL 
prediction (\ref{bfklres}) and the saturation parametrization (\ref{satres}) 
with the new data coming from measurements performed at 
HERA~\cite{zeusnew,h1new}. On one side, our theoretical results are for the 
cross-section (\ref{one}) which is differential with respect to all four 
kinematic variables $x,$ $Q^2,$ $x_J,$ and $k_T.$ On the other side, HERA data 
concern observables which are less differential: $d\sigma/dx,$ $d\sigma/dQ^2,$ 
$d\sigma/dk_T^2,$ and $d\sigma/dxdQ^2dk_T^2.$ Therefore, on the top of the 
Mellin integrations in (\ref{bfklres}) and (\ref{satres}), one has to carry out 
a number of integrations over the kinematic variables which have to be done 
taking into account the kinematic cuts applied by the different experiments. A 
detailed description of how we performed those integrations in given in Appendix 
A and the resulting cross-sections that can be compared to the data are given in 
Appendix B. The method allows for a direct comparison of the data with 
theoretical predictions but it does not allow to control the overall 
normalization. In the following studies, we therefore compare only spectra and 
will not refer to normalizations anymore. As already mentioned, one does not 
adjust any of the parameters of Table I. 

\begin{figure}[t]
\begin{center}
\epsfig{file=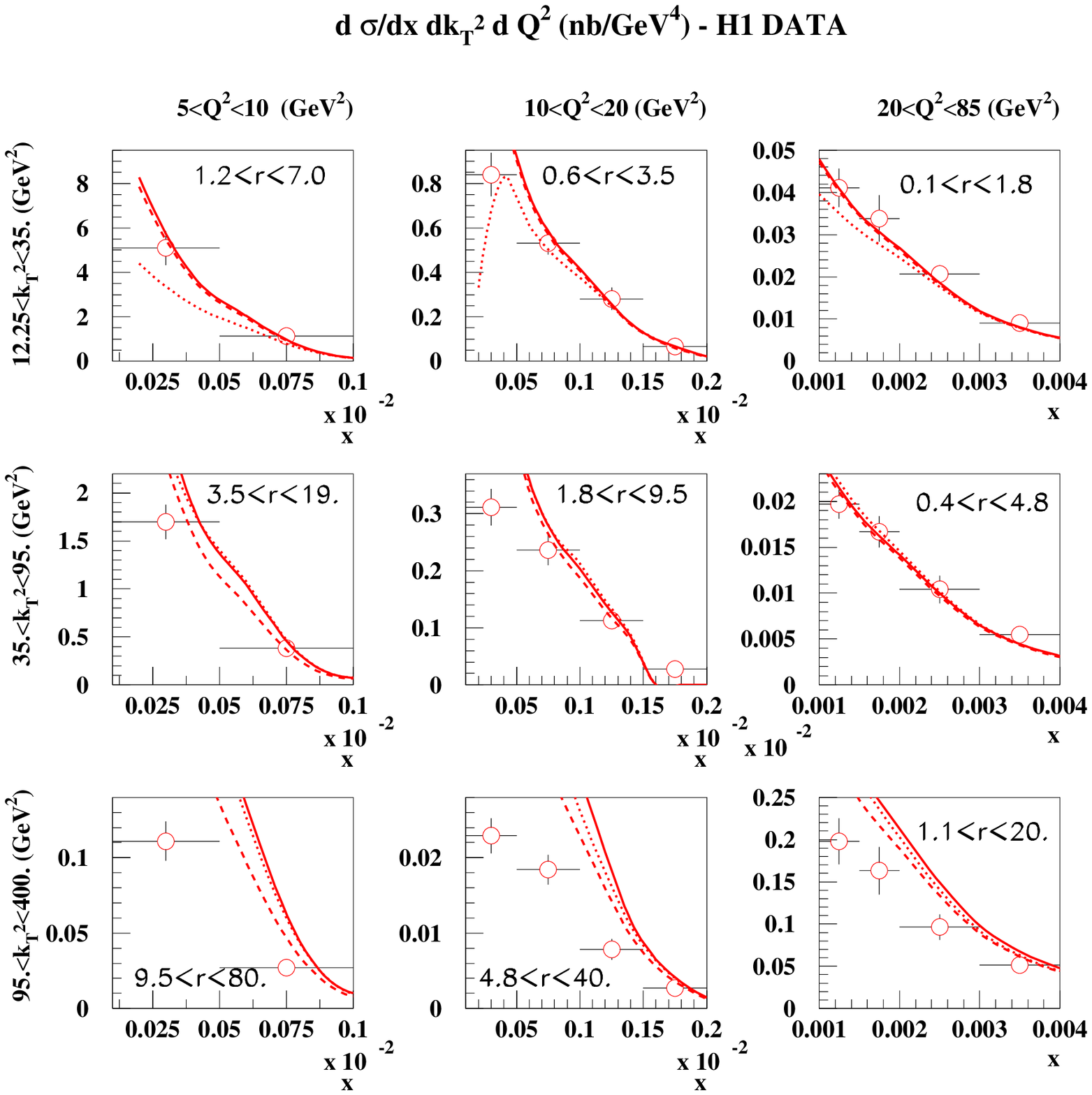,width=14cm}
\caption{The forward-jet cross-section $d\sigma/dxdQ^2dk_T^2.$ The points are 
measurement by the H1 collaboration. The lines are comparisons with BFKL-LL
predictions (full lines) and the two saturation parametrizations (dotted and 
dashed lines for strong and weak saturation respectively). In the regime where 
on expects small$-x$ effects to be important 
($r\!\equiv\!k_T^2/Q^2\!\sim\!1$), there is a good description of the data. In 
the regime where $r\!\gg\!1,$ the small$-x$ parametrizations do not reproduce 
the data as expected from the hierarchy of the hard scales.}
\end{center}
\label{F4}
\end{figure} 

Let us start with the observable $d\sigma/dx$ which has been measured by both 
the H1 and ZEUS collaborations and which now features lower values of $x$ than 
the first measurements. The comparison is displayed in Fig.2, the three 
small$-x$ parametrizations describe very well the data. One cannot really 
distinguish between the three curves, except at small values of $x$ where one 
starts to see a difference: the BFKL curve is above the weak-saturation curve 
which is itself above the strong-saturation curve. However the main conclusion 
is that the data seem to feature the BFKL growth, when going to small values of 
$x.$ For comparison, the fixed-order QCD predictions at NLO computed 
in~\cite{zeusnew,h1new} with the DISENT Monte-Carlo program~\cite{disent} are 
reproduced in Fig.2. At the lowest values of $x,$ they do not reproduce the data 
as they are about a factor 1.5 to 2.5 below depending on the experiment and the 
error bars. Even adding a resolved-photon component to the NLO 
predictions~\cite{h1new} does not pull them within the uncertainties, contrary to 
what happened for the previous data~\cite{respho}. This is an interesting 
difference with the forward-pion case~\cite{basu} for which it seems that no 
higher-order effect other than a NLO resolved-photon contribution is needed.

In Fig.3 are represented the two other single differential cross-sections that 
we shall briefly discuss: $d\sigma/dQ^2$ and $d\sigma/dk_T$ measured by the ZEUS 
collaboration. One can see again that the three small$-x$ parametrizations agree 
well with the data, it is a strong result that one is able to describe the $Q^2$ 
and $k_T$ spectra without any adjustment of the parameters as they were only 
fitted to describe the $x$ dependence.

We shall finally compare our predictions with the triple differential 
cross-section $d\sigma/dxdQ^2dk_T^2$ measured by the H1 collaboration. The 
interesting part of this measurement is that it has been carried out in 9 
different bins of $r\!\equiv\!k_T^2/Q^2$ from $0.1\!<\!r\!<\!1.8$ to 
$9.5\!<\!r\!<\!80.$ This allows to test the limits of our parametrizations which 
are supposed to be valid only when $r\!\sim\!1$ as they do not take into account 
any transverse momentum ordering of the gluons emitted in rapidity between the 
forward jet and the photon. The comparisons with the data are shown on Fig.4 and 
one sees the expected trend. The bins which have $r\!\sim\!1$ are well described 
by the small$-x$ parametrizations while the others are not: for the latter, we 
overshoot the data as the BFKL rise towards small values of $x$ is too steep.  
Interestingly enough, the trend is reversed for QCD predictions at NLO: they 
describe better the data which feature large values of $r.$ These observations 
favor the need of the BFKL resummation to describe the $r\!\sim\!1$ data. The 
large$-r$ bins also exhibit a limitation of the saturation model as one can see 
that the strong saturation parametrization lies above the weak saturation 
parametrization. Such a behavior indicates that the model should not be used 
when $k_T^2\!\gg\!Q^2.$ 

Let us comment further on the two saturation parametrizations. While the BFKL 
formula (\ref{bfklres}) is a QCD prediction as it is computed from Feynman 
diagrams~\cite{bfkl}, the saturation parametrization (\ref{satres}) is a 
phenomenological model. The fact that it describes well the data does not call 
for the same conclusions as in the BFKL case. It only exhibits that, as it is 
the case for a number of observables, data are compatible with saturation 
effects even at energies which do not require them. In other words, the 
forward-jet measurement at the present energies cannot distinguish between 
saturation and BFKL effects; one would start seeing a significant difference at 
higher energies.

It is the purpose of the next section to look for such differences, by studying  
another process similar to forward-jets, namely Mueller-Navelet jets, at LHC 
energies.

%%%%%%%%%%%%%%%%%%%%%%%%%%%%%%%%%%%%%%%%%%%%%%%%%%%%%%%%%%%%%%%%%%
%%%%%%%%%%%%%%%%%%%%%%%%   Section 2   %%%%%%%%%%%%%%%%%%%%%%%%%%%
%%%%%%%%%%%%%%%%%%%%%%%%%%%%%%%%%%%%%%%%%%%%%%%%%%%%%%%%%%%%%%%%%%
\section{Towards the LHC: Mueller-Navelet jets}
\label{3}

Mueller-Navelet jet production in a proton-proton collision is represented in 
Fig.5 with the different kinematic variables. We denote $\sqrt{S}$ the total 
energy of the collision, $k_1$ and $k_2$ the transverse momenta of the two 
forward jets and $x_1$ and $x_2$ their longitudinal fraction of momentum with 
respect to the protons as indicated on the figure. In the following, we 
compute the Mueller-Navelet jet cross-section in the high-energy limit, recall 
the BFKL predictions and formulate our saturation model. We then display 
predictions for observables which can be measured at the LHC.

\subsection{Formulation}

As in the original paper~\cite{mnj}, we consider the cross-section differential 
with respect to $x_1$ and $x_2$ and integrated over the transverse momenta of 
the jets with $k_1\!>\!Q_1$ and $k_2\!>\!Q_2.$ $Q_1$ and $Q_2$ represent then 
experimental $k_T-$cuts. Considering the high energy limit, the QCD 
cross-section for Mueller-Navelet jet production reads~\cite{marq,marpes}:
\be
\f{d\sigma^{pp\!\rightarrow\!JXJ}}{dx_1dx_2}=
\f{\pi^2N_c^2}{64}f_{eff}(x_1,Q_1^2)f_{eff}(x_2,Q_2^2)
\int_0^\infty dr\int_0^\infty d\bar{r}
\ Q_1J_1(Q_1r)Q_2J_1(Q_2\bar{r})\ 
\sigma_{(gg)(gg)}(r,\bar{r},Y)\label{mnjets}\ee
with $Y\!=\!\log(x_1x_2S/Q_1Q_2)$ the rapidity assumed to be very large.
$\sigma_{(gg)(gg)}(r,\bar{r},Y)$ is the cross-section in the collision of two 
$gg$ dipoles of size $r$ and $\bar{r}$ with rapidity total Y. As before
$f_{eff}$ is the effective parton distribution function (\ref{sf}).

Let us comment formula (\ref{mnjets}). As before, each forward jet involves 
perturbative values of transverse momenta and moderate values for $x_1$ and 
$x_2.$ This explains the collinear factorization of the two functions $f_{eff};$ 
here we have taken the factorization scales to be $Q^2_1$ and $Q^2_2.$ The 
remaining hard interaction is between two $gg$ dipoles: as we have seen in the 
previous section, each of them describes a gluon emission at high energies. 
Formula (\ref{mnjets}) expresses the Mueller-Navelet jet observable in terms 
of the cross-section $\sigma_{(gg)(gg)}$ which contains the high-energy QCD 
dynamics. This is the similarity with the forward-jet case: the problem is also 
analogous to the one of onium-onium scattering.

\begin{figure}[t]
\begin{center}
\epsfig{file=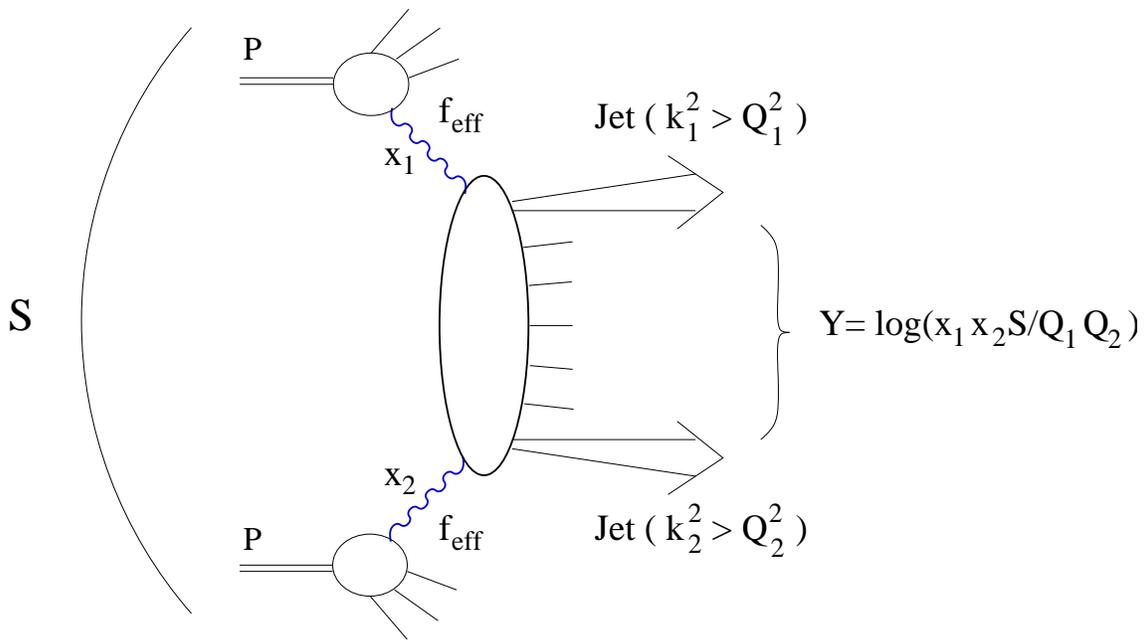,width=15cm}
\caption{Mueller-Navelet jet production in a proton-proton collision. The 
kinematic variables of the problem are displayed. $S$ is the total energy 
squared, $k_1$ and $k_2$ are the transverse momenta of the jets and $x_1$ and 
$x_2$ are their longitudinal momentum fraction with respect to the incident 
protons. $Y$ is the rapidity interval between the hard probes.}
\end{center}
\label{F5}
\end{figure}

Let us first consider the BFKL energy regime, the $gg$-$gg$ dipole-dipole 
cross-section reads
\be
\sigma^{BFKL}_{(gg)(gg)}(r,\bar{r},Y)=\f{2\pi N_c\alpha_s^2}{C_F}\ r^2
\intc{\g}\f{\lr{\bar{r}/r}^{2\g}}{\g^2(1\!-\!\g)^2}\
\exp{\lr{\D\f{\alpha_sN_c}{\pi}\chi(\g)Y}}
\label{sddbis}
\ee
which combined with (\ref{mnjets}) gives
\be
\f{d\sigma^{BFKL}}{dx_1dx_2}=
\f{\pi^3N_c^3\alpha_s^2}{8C_F Q_1^2}f_{eff}(x_1,Q_1^2)f_{eff}(x_2,Q_2^2)
\intc{\g}\f{\lr{Q_1/Q_2}^{2\g}}{\g(1\!-\!\g)}\
\exp{\lr{\D\f{\alpha_sN_c}{\pi}\chi(\g)Y}}\ .
\label{mnjbfkl}\ee
One can easily show that the result is identical to the one obtains using 
$k_T-$factorization~\cite{mnj}. As in the forward-jet case, 
the only undetermined parameter is $\bar\alpha$ which appears in 
the exponential in formula (\ref{mnjbfkl}). We shall consider in this study the 
same value that was used for forward jets, that is $\bar\alpha\!=\!0.16.$

For the saturation parametrization, we use the following $gg$-$gg$ dipole-dipole 
cross-section:
\be
\sigma^{sat}_{(gg)(gg)}(r,\bar{r},Y)=\f{4\pi N_c\alpha_s^2}{C_F}\ \sigma_0
\lr{1-\exp\lr{-\f{r_{\rm eff}^2(r,\bar{r})}{4R_0^2(Y)}}}\ .
\label{sigmaddbis}
\ee
which up to the normalization is the same as (\ref{sigmadd}). The effective 
radius $r_{eff}$ is defined by formula (\ref{reff}) and the saturation radius by 
$R_0(Y)\!=\!e^{-\f{\lambda}2\left(Y-Y_0\right)}/Q_0$ with $Q_0\!\equiv\!1\ GeV.$
Inserting (\ref{sigmaddbis}) into (\ref{mnjets}), one obtains~\cite{marpes}
\bea
\f{d\sigma^{sat}}{dx_1dx_2}=
\f{\pi^3N_c^3\alpha_s^2\sigma_0}{16C_F}f_{eff}(x_1,Q_1^2)f_{eff}(x_2,Q_2^2)
\left\{1-2R_0^2(Y)Q_1Q_2\int_1^\infty 
\f{du}{1+\log(u)}\ I_1\lr{\f{2Q_1Q_2uR_0^2(Y)}{1+\log(u)}}
\right.\nonumber\\\nonumber\\\left.\times
\left[\exp\lr{-\D\f{Q_1^2+u^2Q_2^2}{1+\log(u)}R_0^2(Y)}
+\exp\lr{-\D\f{Q_2^2+u^2Q_1^2}{1+\log(u)}R_0^2(Y)}\right]\right\}\ .
\label{mnjsat}\eea
In the following we consider only the strong saturation parametrization to 
display what could be the maximal expected effects at the LHC. The parameters 
are $\lambda\!=\!0.402$ and $Y_0\!=\!-0.82.$ The normalization $\sigma_0$ is a 
priori not determined but we have fixed it so that at large momenta and small 
$Y,$ one obtains the BFKL result.

\begin{figure}[t]
\begin{center}
\epsfig{file=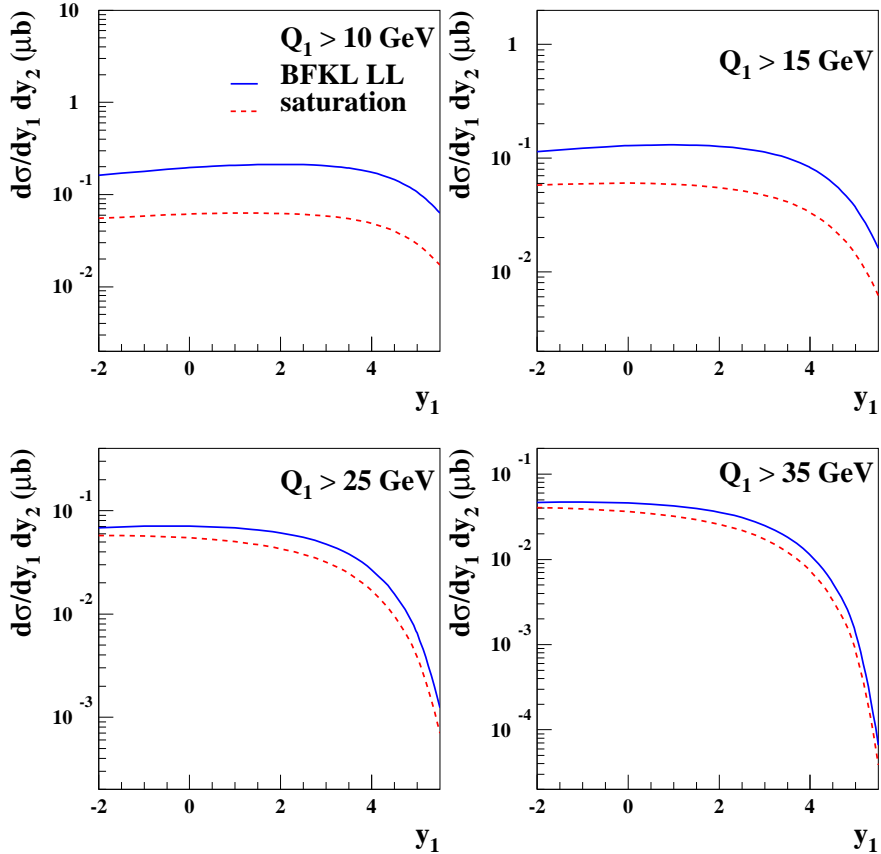,width=12cm}
\caption{The Mueller-Navelet jet cross-section $d\sigma/dy_1dy_2$ as a function 
of $y_1$ for different values of $Q_1.$ The kinematics of the other jet are 
fixed at $Q_2\!=\!30\ \mbox{GeV}$ and $y_2\!=\!-4.5.$ The full lines are BFKL-LL 
predictions and the dashed lines are the saturation 
parametrization.}
\end{center}
\label{F6}
\end{figure} 

\subsection{Phenomenology}

\begin{figure}[t]
\begin{center}
\epsfig{file=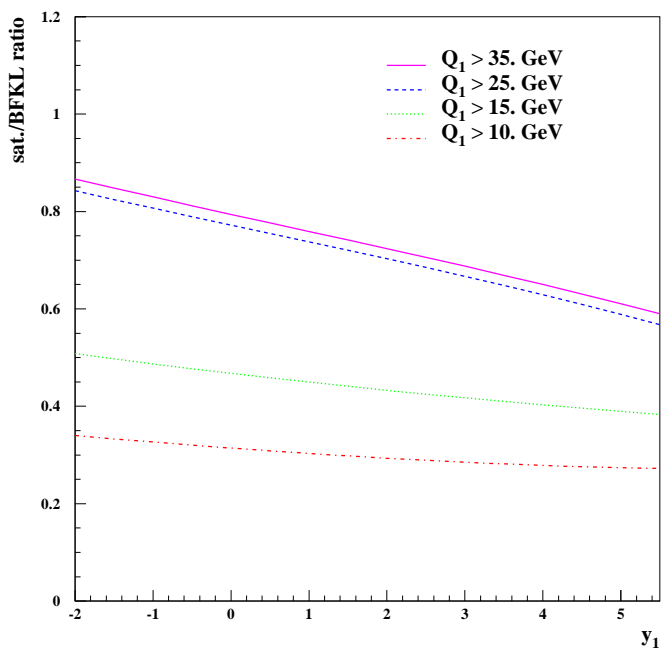,width=9cm}
\caption{Ratio of the saturation and BFKL-LL Mueller-Navelet jet cross-sections 
$d\sigma/dy_1dy_2$ as a function of $y_1$ for different values of $Q_1.$ The 
kinematics of the other jet are fixed at $Q_2\!=\!70\ \mbox{GeV}$ and 
$y_2\!=\!-3.5.$}
\end{center}
\label{F7}
\end{figure} 

We are going to study the dependence of the cross-sections (\ref{mnjbfkl}) and 
(\ref{mnjsat}) as a function of the different kinematic variables $x_1,$ $x_2,$ 
$Q_1,$ and $Q_2.$ We want to consider large rapidities $Y$ which implies very 
forward jets and therefore large values of $x_1$ and $x_2.$ The well-known 
problem is that the cross-section is then damped by the parton distribution 
functions which at large $x$ become very small. This prevents one to see the 
BFKL enhancement of the hard part of the cross-section with rapidity.

A way out of this problem is to consider the following observables ${\cal 
R}_{S/\tilde{S}}:$
\be
{\cal R}_{S/\tilde{S}}\equiv
\f{d\sigma^{pp\!\rightarrow\!JXJ}}{dx_1dx_2}(Q_1,Q_2,S)\Big/
\f{d\sigma^{pp\!\rightarrow\!JXJ}}{dx_1dx_2}(Q_1,Q_2,\tilde{S})\ ,
\label{Rij}
\ee
in other words, cross-section ratios for same jet kinematics and two different 
values of the total energy squared ($S$ and $\tilde{S}$). The advantage of such 
observables is that they are independent of the
parton densities and allow to study more quantitatively the influence of 
small$-x$ effects~\cite{fjets,marpes,mpr}. For instance the BFKL-LL prediction 
is (via a saddle point approximation):
\be
{\cal R}_{S/\tilde{S}}\simeq\lr{\f{S}{\tilde{S}}}^{4\bar\alpha\log(2)}.
\label{rval}\ee
The experimental verification of this at the Tevatron~\cite{mnjtev} was not 
conclusive. The data were found above the prediction (\ref{rval}), however it 
has been argued~\cite{schmidt} that the measurement was biased by the use of 
upper $k_T-$cuts, the choice of equal lower $k_T-$cuts, and hadronization 
corrections. The ratios (\ref{Rij}) also display in a clear way the saturation 
effects~\cite{marpes,mpr} which lead to ratios that, as a function of $Q_1$ and 
$Q_2,$ go from the value (\ref{rval}) to 1 as the momentum cuts decrease into 
the saturation region (see Fig.4 in Ref. \cite{mpr} and Fig.3,4 in Ref. 
\cite{marpes}).

There is however an important experimental limitation to carry out the 
measurement (\ref{Rij}): it would require to run the LHC at two different 
center-of-mass energies. If it turns out not to be possible, then one should 
settle for the cross-section $d\sigma/dx_1dx_2.$ We shall now exhibit some of 
its characteristics, fixing the LHC center-of-mass energy at $\sqrt{S}\!=\!14\ 
\mbox{TeV}.$ Also the absolute normalization is fixed to reproduce the Tevatron 
data at $\sqrt{S}\!=\!1.8\ \mbox{TeV}$ published in~\cite{mnjtev}. These data 
feature somewhat large error bars which leads to a significant uncertainty on 
the normalization for the LHC predictions.

\begin{figure}[t]
\begin{center}
\epsfig{file=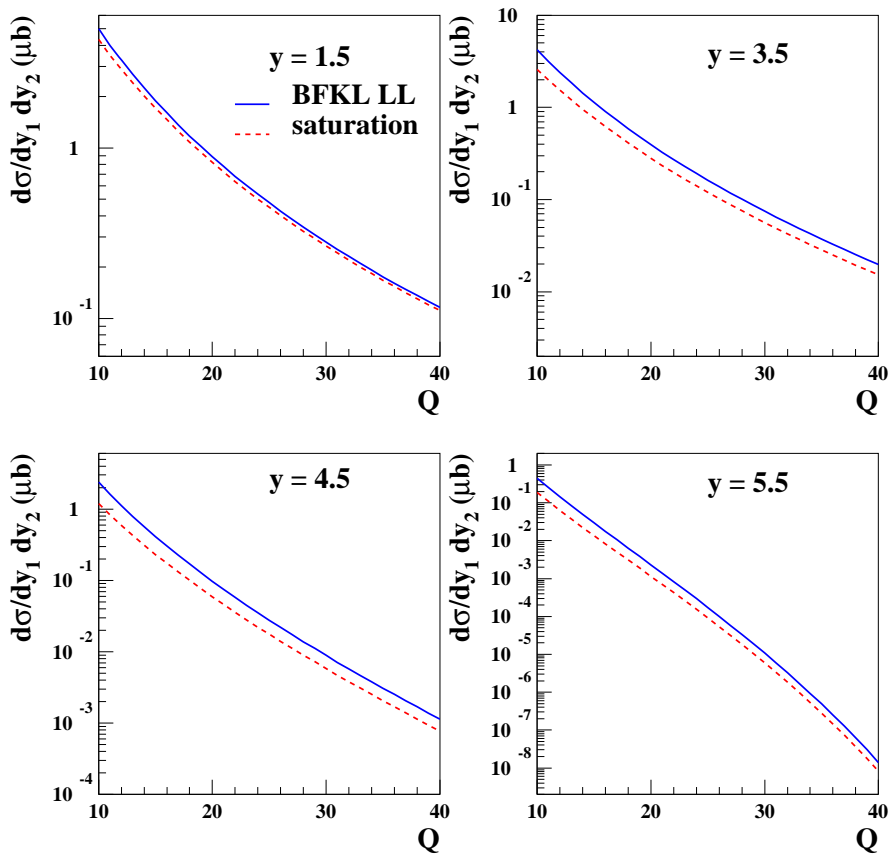,width=12cm}
\caption{The Mueller-Navelet jet cross-section $d\sigma/dy_1dy_2$ as a function 
of $Q\!\equiv\!Q_1\!=\!Q_2$ for different values of $y\!\equiv\!y_1\!=\!-y_2.$ 
The full lines are BFKL-LL predictions and the dashed lines are the saturation 
parametrization.}
\end{center}
\label{F8}
\end{figure} 

Let us introduce the rapidities of the two jets:
\be
y_1=\log\lr{\f{x_1\sqrt{S}}{Q_1}}\ ,\hspace{1cm}
y_2=-\log\lr{\f{x_2\sqrt{S}}{Q_2}}\ .
\label{raps}
\ee
We first considered the case where one of the two jets has fixed kinematics 
$Q_2\!=\!30\ \mbox{GeV}$ and $y_2\!=\!-4.5.$ We looked at the dependence of the 
cross-section $d\sigma/dy_1dy_2\!=\!x_1x_2\ d\sigma/dx_1dx_2$ as a function 
of the other jet kinematic variables. In Fig.6, we plotted the results for the 
BFKL-LL prediction (\ref{mnjbfkl}) and the saturation parametrization 
(\ref{mnjsat}) where the different plots feature the $y_1$ dependence for 
different values of $Q_1.$ As expected, the cross-sections decrease quickly as 
$y_1$ gets large which corresponds to $x_1$ getting closer to one. For each 
value of $Q_1,$ one cannot really see a difference between the behaviors of the 
BFKL and saturation curves as a function of $y_1.$ However the relative 
normalization between the two curves is quite sensitive to the value of $Q_1.$
This is better exhibited on Fig.7 where one displays the ratio of the saturation 
and BFKL results of Fig.6. The ratio goes down to about 0.3 for $Q_1\!=\!10\ 
\mbox{GeV}$ which represents a significant difference between the BFKL and 
saturation predictions. Note that this difference does not appear to be that 
large on Fig.6 where the cross-sections are plotted.

The second case we considered is the symmetric case $Q\!\equiv\!Q_1\!=\!Q_2$ and 
$y\!\equiv\!y_1\!=\!-y_2$ which allows to go to bigger values of $Y.$ We looked 
at the dependence of the cross-section $d\sigma/dy_1dy_2$ as a function of $Q$ 
and $y.$ In Fig.8, we plotted the results for the BFKL prediction 
(\ref{mnjbfkl}) and the saturation parametrization (\ref{mnjsat}) where the 
different plots feature the $Q$ dependence for different values of $y.$ In this 
case, the cross-section falls even faster when $y$ gets big as both $x_1$ and 
$x_2$ get close to 1. Again, because of that, one does not see on the plot the 
difference between the BFKL and the saturation curves, yet it is still quite big 
as shown on Fig.9 where we have displayed the ratio of the saturation and BFKL 
results of Fig.8. For $y\!=\!5.5$ and $Q$ decreasing down to $10\ \mbox{GeV},$ 
the ratio goes down to about 0.4.

\begin{figure}[t]
\begin{center}
\epsfig{file=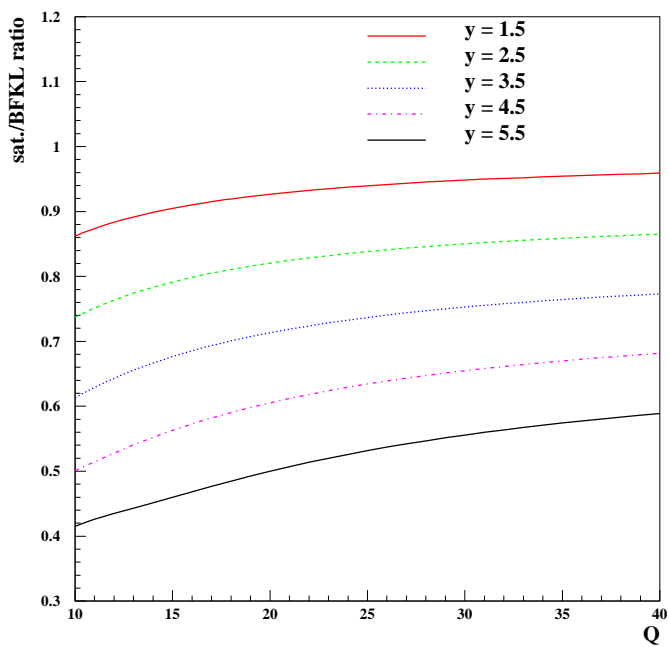,width=9cm}
\caption{Ratio of the saturation and BFKL-LL Mueller-Navelet jet cross-sections 
$d\sigma/dy_1dy_2$ as a function of $Q\!\equiv\!Q_1\!=\!Q_2$ for different 
values of $y\!\equiv\!y_1\!=\!-y_2.$}
\end{center}
\label{F9}
\end{figure} 

We did not include in this study the weak saturation parametrization, the 
corresponding curves would lie in between the BFKL and strong saturation curves 
which are displayed, and even is closer to the BFKL curve. There is a number of 
other plots one could study showing other dependences of $d\sigma/dy_1dy_2$ but 
they are not needed for drawing our conclusions: testing BFKL effects and 
saturation effects with the observable $d\sigma/dy_1dy_2$ at the LHC will be a 
major experimental challenge as one will have to measure cross-sections with a 
great precision. We insist that this is due to the fact that the parton 
distribution functions at large $x$ really damp the cross-section. Obtaining a 
high accuracy is not unfeasible because of the high luminosity at the LHC but 
this will require a very good understanding of the systematics errors. However 
we would like to emphasize the fact that better tests of small$-x$ effects could 
be realized with the measurement of the ratio ${\cal R}_{S/\tilde{S}},$ see 
formula (\ref{Rij}).

%%%%%%%%%%%%%%%%%%%%%%%%%%%%%%%%%%%%%%%%%%%%%%%%%%%%%%%%%%%%%%%%%%
%%%%%%%%%%%%%%%%%%%%%%%   Conclusion   %%%%%%%%%%%%%%%%%%%%%%%%%%%
%%%%%%%%%%%%%%%%%%%%%%%%%%%%%%%%%%%%%%%%%%%%%%%%%%%%%%%%%%%%%%%%%%
\section{Conclusions}
\label{6}

Let us summarize the main results of the paper. The first part of the work 
was devoted to the study of the forward-jet measurement. We started by 
computing the QCD cross-section for forward-jet production (\ref{one}) in the 
high-energy (small$-x$) limit. We recalled the BFKL-LL predictions 
(\ref{bfklres}) and also formulated the phenomenological model (\ref{satres}) 
that takes into account saturation effects. We then compared the BFKL and 
saturation-model predictions to the recent data from HERA for a number of 
observables: $d\sigma/dx,$ $d\sigma/dQ^2,$ $d\sigma/dk_T^2,$ and 
$d\sigma/dxdQ^2dk_T^2.$ 

We obtained a very good agreement with the BFKL predictions and saturation 
parametrizations also show compatibility with the data. Along with the fact that 
QCD at NLO predictions do not reproduce the small$-x$ data, this observation 
leads us to the conclusion that the present forward-jet data display the BFKL 
enhancement when going to small values of $x.$ 

However, to make a definitive statement, one would have to make comparisons with 
BFKL predictions at next-to-leading-logarithmic (NLL) accuracy. The latter are 
under investigations and will hopefully be available soon. In the mean time, let 
us discuss the expected qualitative impact of these BFKL-NLL corrections. Because 
we are describing a kinematic regime in which $k_T^2\!\sim\!Q^2,$ one can infer 
that they should be small. For instance, the contribution coming from the running 
of the coupling between those two scales would be unimportant. By comparison, the 
BKFL-NNL corrections seem to be very large for the proton structure function 
measurement~\cite{nll}, in which case one easily understands why: the evolution 
takes place in a large range, from the soft proton scale up to the hard scale 
$Q^2.$ The situation is much different for forward jets.

In the second part of the paper, we investigated small$-x$ effects for 
Mueller-Navelet jets in the LHC energy range, using the parameters that 
successfully describe forward-jets at HERA. We compared the BFKL-LL predictions 
(\ref{mnjbfkl}) with those of the saturation model (\ref{mnjsat}) and concluded 
that the measurement of the simple cross-section will require a great precision 
to test the different scenarios. We argued that a better option to look for 
small$-x$ effects was to measure the ratio of cross-sections (\ref{Rij}) which 
implies running the LHC at two different energies.

On longer time scales, the international linear collider would give the 
opportunity to measure the virtual photon-virtual photon total cross-section at 
very high energies. This would also offer great 
possibilities~\cite{gambfkl,gamsat} for testing the BFKL enhancement and the 
saturation regime of QCD.

%%%%%%%%%%%%%%%%%%%%%%%%%%%%%%%%%%%%%%%%%%%%%%%%%%%%%%%%%%%%%%%%%%
%%%%%%%%%%%%%%%%%%%%%   Acknowledgments   %%%%%%%%%%%%%%%%%%%%%%%%
%%%%%%%%%%%%%%%%%%%%%%%%%%%%%%%%%%%%%%%%%%%%%%%%%%%%%%%%%%%%%%%%%%
\begin{acknowledgments}

The authors would like to thank Robi Peschanski for useful
comments and fruitful discussions.
 
\end{acknowledgments}
%%%%%%%%%%%%%%%%%%%%%%%%%%%%%%%%%%%%%%%%%%%%%%%%%%%%%%%%%%%%%%%%%%
%%%%%%%%%%%%%%%%%%%%%%%%%  Appendix  %%%%%%%%%%%%%%%%%%%%%%%%%%%%%
%%%%%%%%%%%%%%%%%%%%%%%%%%%%%%%%%%%%%%%%%%%%%%%%%%%%%%%%%%%%%%%%%%
\section*{Appendix A: on the integration method}

To compare the forward-jet cross-section (\ref{one}) obtained from the BFKL-LL 
prediction (\ref{bfklres}) and the saturation parametrization (\ref{satres}) 
with the data for observables which are less differential ($d\sigma/dx,$ 
$d\sigma/dQ^2,$ $d\sigma/dk_T^2,$ and $d\sigma/dxdQ^2dk_T^2$), one has to carry 
out a number of integrations over the kinematic variables. They have to be done 
while properly taking into account the kinematic cuts applied by the different 
experiments. This Appendix deals with these issues.

Let us start from the quadruple differential cross-section 
$d\sigma/dxdQ^2dx_Jdk_T^2,$ see formula (\ref{one}). First one performs the 
Mellin integrations of (\ref{bfklres}) and (\ref{satres}). Then we choose the 
appropriate variables for the remaining integrations: to avoid 
numerical problems in the integral calculations, we chose variables which 
lead to the weakest possible dependence of the differential cross-section. We 
noticed that the best choice is $1/Q^2,$ $1/k_T^2,$ $\log(1/x_J),$ and 
$\log(1/x).$
Since the experimental measurements are not differential with respect to $x_J,$ 
we carry out the integration 
\be
\f{d^{(3)}\sigma}{dxdQ^2dk_T^2}=\int \f{x_J\ d^{(4)}\sigma}{dxdQ^2dx_Jdk_T^2}\ 
d\log\lr{\f1{x_J}}\ .\nonumber\ee

With the BFKL-LL formula (\ref{bfklres}), the convergence is fast enough so that 
one can perform all the remaining integrations to obtain any of the four 
observables mentioned above. With the saturation formula (\ref{satres}), because 
of the extra Mellin integration and the time it takes to compute the 
functions $\Psi,$ performing all the remaining integrations would require 
important numerical work. We chose to use another method to obtain the 
cross-sections in that case. We shall describe it now with the example of 
$d\sigma/dx.$

For a given value of $x,$ the first step is to compute the differential 
cross-section 
\be
\f{d\sigma^{BFKL}}{dx}=\int Q^4 k_T^4\ \f{d^{(3)}\sigma^{BFKL}}{dxdQ^2dk_T^2}\ 
d\lr{\f1{Q^2}}\ d\lr{\f1{k_T^2}}\ \nonumber\ee
for the BFKL case. The second step is to compute the bin center ($k_{TC}^2$, 
$Q_C^2$) defined as follows:
\be
\f{d^{(3)}\sigma^{BFKL}}{dxdQ^2dk_T^2}(Q_C^2,k_{TC}^2)\equiv 
\f{d\sigma^{BFKL}/dx}{\int dQ^2 dk_T^2}\ .
\nonumber\ee
The bin center is thus the point in the $(k_T^2, Q^2)$ phase space where the
differential cross-section in $k_T^2$ and $Q^2$ is equal to the integral
over the bin divided by the bin size (we will specify the integration limits 
later on). The third step is to obtain the cross 
section for the saturation case. We compute the cross-section at the bin center 
($k_{TC}^2$, $Q_C^2$):
\be
\f{d\sigma^{sat}}{dx}=\f{d^{(3)}\sigma^{sat}}{dxdQ^2dk_T^2}(Q_C^2,k_{TC}^2)\ 
\int dQ^2 dk_T^2\ .\nonumber\ee
This procedure is valid if the bin center does not change much between the BFKL 
and saturated cross-sections. In other words, it means that the difference 
between the BFKL and saturated cross-sections is small. We saw in Section II-E 
that this is indeed the case. The method is easily adapted to the case 
of $d\sigma/dQ^2$ for which one finds a bin center $(x_C,k_{TC}^2)$ for each 
value of $Q^2$ and to the case of $d\sigma/dk_T^2$ with a bin center 
$(x_C,Q_C^2)$ for each value of $k_T.$

For the triple differential cross-section $d\sigma/dxdQ^2dk_T^2$ which is 
measured as a function of $x,$ integrating over $x_J$ is not enough since one 
does not know the $(Q^2,k_T^2)$ bin-center. Instead, for a given value of $x,$ 
one integrates also over $Q^2$ and $k_T^2$ and then divide the result by the 
$Q^2$ and $k_T^2$ bin sizes to obtain $d\sigma/dxdQ^2dk_T^2.$ This is done for 
the BFKL case and one uses again the method described above to compute the 
cross-section in the saturation case.

The other difficulty arises when setting the integration limits as one has to 
take into account the correlations between the kinematic variables (for 
instance $y\!=\!Q^2/sx\!<\!1$) and the cuts applied by the experiments (for 
instance cuts on the forward jet phase space). There are two ways to take these 
into account: either appropriately set the limits of integration while computing 
the integrals or evaluate later the phase space correction due to the 
experimental cuts. We are going to use both, since it is not possible to include 
all experimental cuts while computing the integrals.

Table II is a list of the different set of cuts used by the H1 and ZEUS 
experiments to carry out their measurements. For the ZEUS cuts, we only consider 
what they call the ``forward-BFKL phase space'' which corresponds the most to 
the Regge limit kinematics. Refering to this table, let us enumerate the 
integration limits which are used for the different integral calculations:
\begin{itemize}
\item {\bf $d \sigma / dx$  for H1:} We integrate over $1/k_T^2$ with the 
limit on $k_T^2$ defined by $0.5\!<\!k_T^2/Q^2\!<\!5$ (this is an extra cut that 
H1 applies to this measurement only), over $1/Q^2$ with 
$Q^2 \!<\!sx$, and over $\log(1/x_J)$ with $0.035\!<\!x_J\!<1$
\item {\bf $d \sigma / dx$ for ZEUS:} We integrate over $1/k_T^2$ with the 
limit on $k_T^2$ defined by $0.5\!<\!k_T^2/Q^2\!<\!2,$ over $1/Q^2$ with $Q^2 
\!<\!sx$, and over $\log(1/x_J)$ with $2\!<\!\log(1/x_J)\!<3$
\item {\bf $d \sigma / dQ^2$ for ZEUS:}  We integrate over $1/k_T^2$ with the 
limit on $k_T^2$ defined by $0.5\!<\!k_T^2/Q^2\!<\!2,$ over $\log(1/x)$ with $x 
\!\ge \!Q^2 / s$, and over $\log(1/x_J)$ between 2 and 3 
\item {\bf $d \sigma / dk_T^2$ for ZEUS:} We integrate over $\log(1/x)$ with 
$x\!\ge\!Q^2/s,$ over $1/Q^2$ with the limits on $Q^2$ defined by 
$0.5\!<\!k_T^2/Q^2\!<\!2,$ and over $\log(1/x_J)$ between 2 and 3
\item {\bf $d\sigma/dxdQ^2dk_T^2$ for H1:} The $1/k_T^2$ and $1/Q^2$ limits of 
the integrals are defined by the bin values measured by the H1 collaboration 
with also the kinematic constraint $ 0.1\!<\!y = Q^2 / s x \!<\!0.7.$ The 
$\log(1/x_J)$ limits are obtained taking into account the cuts on the 
forward-jet angle which leads to $ 1.7354 \!<\!\log(1/x_J) \!<\!2.7942.$
\end{itemize}

\begin{table}[t]
\begin{center}
\begin{tabular}{|c||c|} \hline
 H1 & ZEUS \\ 
\hline\hline
$E_e \ge 10$ GeV &  $E_e \ge 10$ GeV \\
$0.1 \le y \le 0.7$ &  $0.04 \le y \le 1.$  \\ 
$10^{-4} \le x \le 4.10^{-3}$ &    \\
$ 5 < Q^2 < 85$ GeV$^2$ &  $Q^2 > 25$ GeV$^2$  \\
$k_T > 3.5$ GeV & $k_T > 6$ GeV  \\
$7 \le \theta_J \le 20$ degrees &  $2 < \eta_J < 3$ \\
$x_J > 0.035$   &     \\ 
 & $0.5 <k_T^2 / Q^2 < 2$ \\
\hline
\end{tabular}
\end{center}\caption{ZEUS (``forward-BFKL phase space'')~\cite{zeusnew} and 
H1~\cite{h1new} cuts to define the forward-jet phase space. $E_e$ is the energy 
of the outgoing electron and $\eta_J\!=\!\log(1/x_J)\!=\!-\log\tan(\theta_J/2).$ 
The other kinematic variables have been defined in the text.}
\end{table}

The effects of the cuts defined in Table II which are not used above need
to be computed using a toy Monte Carlo. They are modeled by bin-per-bin {\it 
correction factors} that multiply the cross-sections obtained as described 
above. This is how one proceeds: we generate flat distributions in the variables 
$1/k_T^2,$ $1/ Q^2,$ $\log(1/x_J),$ and $\log(1/x)$ using reference intervals 
which include the whole experimental phase-space (the azimuthal angle of the jet 
is not used in the generation since all the cross-section measurements are 
independent of that angle). In practice, we get the correction factors by 
counting the numbers of events which fullfil the experimental cuts for each 
$x-$bin when we compute $d\sigma/dx$, each $Q^2-$bin when we compute 
$d\sigma/dQ^2$ and so on. The correction factors are obtained by the ratio of 
the number of events which pass the experimental cuts and the kinematic 
constraints to the number of events which fullfil only the kinematic 
constraints, i.e. the so-called reference bin. Of course the experimental or 
kinematic cuts which have been applied already while computing the integrals are 
not applied in this study to avoid double counting effects.

This method allows for a direct comparison of the data with theoretical 
predictions but it does not allow to control the overall normalization. This 
would require a full Monte-Carlo. Note that we did not use one in order to avoid 
any strong model dependence of the correction factors as they are only due to 
kinematic-cut effects. The derivation of the correction factors is independent 
of the theoretical input. They are given in Appendix B and they can be used to 
test any model suitable for the forward-jet cross-section, providing the same 
integration method as described above is used.

\section*{Appendix B: tables with correction factors and resulting 
cross-sections}

In this section, we list the corrections factors that we obtained for the 
observables $d\sigma/dx$ (H1 and ZEUS), $d\sigma/dQ^2$ (ZEUS), $d\sigma/dk_T^2$ 
(ZEUS), and $d\sigma/dxdQ^2dk_T^2$ (H1). We also give the resulting 
cross-sections for the different points that we used to draw the curves on 
Fig.2, Fig.3, and Fig.4.

\vspace{0.5cm}

\begin{table}[h]
%\scriptsize
\begin{minipage}[t]{88mm}

\begin{center}
\begin{tabular}{|c||c||c|c|c|} \hline
 x & factor & bfkl-ll & weak sat. & strong sat. \\ 
\hline\hline
0.00015 &  0.24 &  1200.  &  1046.  &  897. \\ 
0.0005  &  0.81 &   805.  &   785.  &  722. \\ 
0.0010  &  0.86 &   371.  &   365.  &  360. \\ 
0.0015  &  0.80 &   205.  &   202.  &  203. \\ 
0.0020  &  0.66 &   117.  &   114.  &  116. \\ 
0.0025  &  0.54 &    72.0 &    70.4 &   71.7 \\ 
0.0030  &  0.45 &    47.1 &    45.9 &   46.9 \\ 
0.0035  &  0.38 &    32.3 &    31.4 &   32.1 \\ 
0.0040  &  0.31 &    22.5 &    21.8 &   22.3 \\ 
\hline
\end{tabular}
\end{center}

\end{minipage}
\hspace{\fill}
\begin{minipage}[t]{88mm}

\begin{center}

\begin{tabular}{|c||c||c|c|c|} \hline
 x & factor & bfkl-ll & weak sat. & strong sat. \\ 
\hline\hline
0.00075 &  0.13 &  39.3  &  34.5  &  31.1 \\ 
0.0017  &  0.43 &   62.5  &   58.1  &  56.0 \\ 
0.004  &  0.48 &   25.4  &   24.7  &  24.6 \\ 
0.01  &  0.34 &   3.33  &   3.38  &  3.34 \\ 
0.025  &  0.13 &   0.106  &   0.109  &  0.106 \\ 
\hline
\end{tabular}
\end{center}

\end{minipage}
\caption{Correction factors due to experimental cuts and the resulting corrected 
cross-sections for $d \sigma /dx$ in nb for BFKL-LL, weak saturation and strong 
saturation (see Fig.2). Left Table: for H1 cuts, right Table: for ZEUS cuts.}
\end{table}

\begin{table}[h]
%\scriptsize
\begin{minipage}[t]{88mm}

\begin{center}
\begin{tabular}{|c||c||c|c|c|} \hline
 $Q^2$ (GeV${}^2$) & factor & bfkl-ll & weak sat. & strong sat. \\ 
\hline\hline
35. &  0.60 &  6.75  &  5.36  &  3.81 \\ 
65.  &  0.70 &   1.40  &   1.03  &  0.843 \\ 
150.  &  0.88 &   0.184  &   0.139  &  0.137 \\ 
350.  &  0.88 &   0.0127  &   0.0112  &  0.0112 \\ 
950.  &  0.96 &   0.000611  &   0.000549  &  0.000549 \\ 
\hline
\end{tabular}
\end{center}

\end{minipage}
\hspace{\fill}
\begin{minipage}[t]{88mm}

\begin{center}
\begin{tabular}{|c||c||c|c|c|} \hline
 $k_T$ (GeV) & factor & bfkl-ll & weak sat. & strong sat. \\ 
\hline\hline
7. &  0.32 &  71.9  &  72.4  &  71.6 \\ 
9.  &  0.31 &   21.1  &   20.2  &  20.4 \\ 
12.  &  0.28 &   6.23  &   5.59  &  5.63 \\ 
17.5  &  0.22 &   1.01  &   0.828  &  0.828 \\ 
25.  &  0.15 &   0.119  &   0.0875  &  0.0875 \\ 
\hline
\end{tabular}
\end{center}

\end{minipage}
\caption{Correction factors due to experimental cuts and the resulting corrected 
cross-sections for BFKL-LL, weak saturation and strong saturation. Left Table: 
for $d \sigma /dQ^2$ in pb/GeV$^2$, right Table: for $d \sigma /dk_T$ in pb/GeV 
(see Fig.3).}
\end{table}

\begin{table}[ht]
%\scriptsize
\footnotesize
\begin{center}
\begin{tabular}{|c|c|c||c||c|c|c|} \hline
 $k_T^2$ (GeV${}^2$) & $Q^2$ (GeV${}^2$) & x & factor & bfkl-ll & weak sat. & 
strong sat. \\ 
\hline\hline
$12.25<k_T^2<35$  &  $5<Q^2<10$  & 0.0002  &  0.065 &  8.28   &  7.85   & 4.40  
\\ 
 & &  0.0004   &  0.065   &  3.86   &  3.69    & 2.53 \\ 
 & &  0.0006   &  0.065   &  2.04   &  1.96    &  1.51 \\ 
 & &  0.0008   &  0.065   &  0.705   & 0.674     &  0.591 \\ 
 & &  0.001    &  0.065   &  0.160   &  0.151    & 0.151 \\ \hline
$12.25<k_T^2<35$  &  $10<Q^2<20$  & 0.0002  &  0.042 &  0.571   &  0.551   &
0.332  \\ 
 & &  0.0004   &  0.065   &  1.15   &  1.12    & 0.826 \\ 
 & &  0.0006   &  0.065   &  0.752   & 0.735     & 0.612 \\ 
 & &  0.0008   &  0.065   &  0.544   &  0.530    & 0.473 \\ 
 & &  0.001    &  0.065   &  0.416   & 0.406     & 0.376 \\ 
 & &  0.0012   &  0.065   &  0.289   & 0.280     & 0.275 \\ 
 & &  0.0014   &  0.065   &  0.171   & 0.166     & 0.169 \\ 
 & &  0.0016   &  0.065   &  9.90e-2   & 9.53e-2     & 9.95e-2 \\ 
 & &  0.0018   &  0.065   &  5.70e-2   & 5.07e-2     & 5.34e-2 \\ 
 & &  0.002    &  0.065   &  2.20e-2   & 2.11e-2     & 2.25e-2 \\ \hline
$12.25<k_T^2<35$  &  $20<Q^2<85$  & 0.001  &  0.060 &  4.79e-2   &  4.72e-2   &
3.97e-2  \\ 
 & &  0.0015   &  0.060   &   3.50e-2  &  3.45e-2     & 3.09e-2 \\ 
 & &  0.002    &  0.063   &   2.68e-2  &  2.63e-2     &	2.46e-2 \\ 
 & &  0.0025   &  0.065   &   1.86e-2  &  1.82e-2     & 1.76e-2 \\ 
 & &  0.003    &  0.065   &   1.20e-2  &  1.17e-2     & 1.15e-2 \\ 
 & &  0.0035   &  0.065   &   8.08e-3  &  7.90e-3     & 7.90e-3 \\ 
 & &  0.004    &  0.065   &   5.60e-3  &  5.47e-3     & 5.52e-3 \\ \hline
\end{tabular}
\end{center}\caption{Correction factors due to experimental cuts and the 
resulting corrected cross-sections for $d \sigma /dxdk_T^2dQ^2 $ in nb/GeV$^4$ 
(bins with $12.25\!<\!k_T^2\!<\!35\ \mbox{GeV}^2$) for BFKL-LL, weak saturation 
and strong 
saturation (see Fig.4).}
\end{table}

\begin{table}[ht!]
%\scriptsize
\footnotesize
\begin{center}
\begin{tabular}{|c|c|c||c||c|c|c|} \hline
 $k_T^2$ (GeV${}^2$) & $Q^2$ (GeV${}^2$) & x & factor & bfkl-ll & weak sat. & 
strong sat. \\ 
\hline\hline
$35<k_T^2<95$  &  $5<Q^2<10$  & 0.0002  &  0.22 &  3.93   & 3.35    & 3.28  \\ 
 & &  0.0004   &  0.22   & 1.86    &  1.57     &  1.82 \\ 
 & &  0.0006   &  0.22   & 1.05    &  0.836     &  1.08 \\ 
 & &  0.0008   &  0.22   & 0.314    &  0.250     & 0.321\\ 
 & &  0.001    &  0.22   & 7.41e-2    &  6.72e-2     & 7.58e-2 \\ \hline
$35<k_T^2<95$  &  $10<Q^2<20$  & 0.0002  &  0.14 &  0.272   &  0.250   & 0.240 
 \\ 
 & &  0.0004   &  0.22   &  0.551   &  0.519     & 0.521 \\ 
 & &  0.0006   &  0.22   &  0.360   &  0.340     & 0.355 \\ 
 & &  0.0008   &  0.22   &  0.261   &  0.245     & 0.263 \\ 
 & &  0.001    &  0.22   &  0.204   &  0.188     & 0.213  \\ 
 & &  0.0012   &  0.22   &  0.142   &  0.131     & 0.149 \\ 
 & &  0.0014   &  0.22   &  8.40e-2   & 7.80e-2      & 8.79e-2 \\ 
 & &  0.0016   &  0.22   &  4.82e-2   & 4.52e-2      & 5.04e-2 \\ 
 & &  0.0018   &  0.22   &  2.57e-2   & 2.42e-2      & 2.68e-2 \\ 
 & &  0.002    &  0.22   &  1.08e-2   & 1.02e-2      & 1.12e-2 \\ \hline
$35<k_T^2<95$  &  $20<Q^2<85$  & 0.001  &  0.20 & 2.49e-2    & 2.44e-2    &
2.55e-2  \\ 
 & &  0.0015   &  0.20   &  1.85e-2   &   1.81e-2    & 1.91e-2 \\ 
 & &  0.002    &  0.21   &  1.42e-2   &   1.38e-2    & 1.46e-2 \\ 
 & &  0.0025   &  0.22   &  9.97e-3   &   9.67e-3    & 1.02e-2 \\ 
 & &  0.003    &  0.22   &  6.53e-3   &    6.33e-3   & 6.63e-3 \\ 
 & &  0.0035   &  0.22   &  4.47e-3   &   4.34e-3    & 4.47e-3 \\ 
 & &  0.004    &  0.22   &  3.15e-3   &   3.04e-3    & 3.13e-3 \\ \hline
\end{tabular}
\end{center}\caption{Correction factors due to experimental cuts and the 
resulting corrected cross-sections for $d \sigma /dxdk_T^2dQ^2 $ in nb/GeV$^4$ 
(bins with $35\!<\!k_T^2\!<\!95\ \mbox{GeV}^2$) for BFKL-LL, weak saturation and 
strong 
saturation (see Fig.4).}
\end{table}

\begin{table}[ht]
%\scriptsize
\footnotesize
\begin{center}
\begin{tabular}{|c|c|c||c||c|c|c|} \hline
 $k_T^2$ (GeV${}^2$) & $Q^2$ (GeV${}^2$) & x & factor & bfkl-ll & weak sat. & 
strong sat. \\ 
\hline\hline
$95<k_T^2<400$  &  $5<Q^2<10$  & 0.0002  &  0.31 &  0.539   & 0.384    & 
0.540 \\ 
 & &  0.0004   &  0.31   &  0.253   &   0.181    & 0.246 \\ 
 & &  0.0006   &  0.31   &  0.135   &    9.67e-2   &  0.127 \\ 
 & &  0.0008   &  0.31   &  4.47e-2   &   3.43e-2    & 4.35e-2 \\ 
 & &  0.001    &  0.31   &  1.02e-2   &  7.87e-3     & 9.62e-3 \\ \hline
$95<k_T^2<400$  &  $10<Q^2<20$  & 0.0002  &  0.20 &  4.28e-2   &  2.99e-2   &
4.00e-2   \\ 
 & &  0.0004   &  0.31   &  7.71e-2   &  6.34e-2     & 8.01e-2 \\ 
 & &  0.0006   &  0.31   &  5.04e-2   &  4.14e-2     & 5.13e-2 \\ 
 & &  0.0008   &  0.31   &  3.78e-2   &  2.99e-2     & 3.59e-2 \\ 
 & &  0.001    &  0.31   &  3.07e-2   &  2.28e-2     & 2.62e-2 \\ 
 & &  0.0012   &  0.31   &  2.08e-2   &  1.60e-2     & 1.82e-2 \\ 
 & &  0.0014   &  0.31   &  1.17e-2   &  9.62e-3     & 1.09e-2 \\ 
 & &  0.0016   &  0.31   &  6.55e-3   &  5.61e-3     & 6.38e-3 \\ 
 & &  0.0018   &  0.31   &  3.49e-3   &  3.03e-3     & 3.40e-3 \\ 
 & &  0.002    &  0.31   &  1.48e-3   &  1.28e-3     & 1.42e-3 \\ \hline 
$95<k_T^2<400$  &  $20<Q^2<85$  & 0.001  &  0.29 &  3.65e-3   &  3.28e-3   &
3.69e-3  \\ 
 & &  0.0015   &  0.29   &  2.78e-3   &  2.47e-3     & 2.69e-3 \\ 
 & &  0.002    &  0.30   &  2.12e-3   &  1.89e-3     & 2.03e-3 \\ 
 & &  0.0025   &  0.31   &  1.49e-3   &  1.33e-3     & 1.41e-3  \\ 
 & &  0.003    &  0.31   &  9,84e-4   &  8.86e-4     & 9.20e-4 \\ 
 & &  0.0035   &  0.31   &  6.76e-4   &  6.13e-4     & 6.30e-4 \\ 
 & &  0.004    &  0.31   &  4.79e-4   &  4.36e-4     & 4.44e-4 \\ \hline
\end{tabular}
\end{center}\caption{Correction factors due to experimental cuts and the 
resulting corrected cross-sections for $d \sigma /dxdk_T^2dQ^2 $ in nb/GeV$^4$ 
(bins with $95\!<\!k_T^2\!<\!400\ \mbox{GeV}^2$) for BFKL-LL, weak saturation 
and strong 
saturation (see Fig.4).}
\end{table}

%%%%%%%%%%%%%%%%%%%%%%%%%%%%%%%%%%%%%%%%%%%%%%%%%%%%%%%%%%%%%%%%%%
%%%%%%%%%%%%%%%%%%%%%%   Bibliography   %%%%%%%%%%%%%%%%%%%%%%%%%%
%%%%%%%%%%%%%%%%%%%%%%%%%%%%%%%%%%%%%%%%%%%%%%%%%%%%%%%%%%%%%%%%%%

\end{document}